%% file: report.tex
\documentclass[a4paper, 10pt]{article}

\usepackage[numbers,sort&compress]{natbib}
\usepackage{changes}
\usepackage{geometry}
\usepackage{textcomp}
\usepackage[T2A]{fontenc}
\usepackage[utf8]{inputenc}
\usepackage{bold-extra}
\usepackage{amsmath,amssymb,amsfonts}
\usepackage{algorithmic}
\usepackage{graphicx}
\usepackage{tabularx}
\usepackage{subcaption}
\usepackage{textcomp}
\usepackage{xcolor}
\usepackage{accents}
\usepackage{footnote}
\usepackage[flushleft]{threeparttable}
\usepackage{booktabs}
\usepackage{wasysym}
\usepackage{hyphenat}
\usepackage{parskip}
\usepackage{authblk}
\usepackage{relsize}
\renewcommand{\textcurrency}{\currency}

\DeclareMathOperator{\sgn}{sgn}

\usepackage[colorinlistoftodos,prependcaption,textsize=footnotesize]{todonotes}

\title{Performance-Feedback Autoscaling with Budget Constraints for Cloud-based Workloads of Workflows\\[1em]\relsize{-1.5}Technical Report}

\author[1]{Alexey Ilyushkin}
\author[2]{Andr\'e Bauer}
\author[3]{Alessandro V. Papadopoulos}
\author[4]{\\Ewa Deelman}
\author[5,1]{Alexandru Iosup}
\affil[1]{Delft University of Technology, the Netherlands}
\affil[2]{University of W\"urzburg, Germany}
\affil[3]{M\"alardalen University, Sweden}
\affil[4]{University of Southern California, USA}
\affil[5]{Vrije Universiteit Amsterdam, the Netherlands}

\date{\vspace{-5ex}}

\begin{document}

\topskip0pt
\maketitle

\vspace{0.5em}
\begin{center}
a.s.ilyushkin@tudelft.nl, andre.bauer@uni-wuerzburg.de, \\ alessandro.papadopoulos@mdh.se, deelman@isi.edu, a.iosup@vu.nl
\end{center}

\thispagestyle{empty}
\vspace{5em}
\begin{abstract}
The growing popularity of workflows in the cloud domain
promoted the development of sophisticated autoscaling policies
that allow automatic allocation and deallocation of resources.
However, many state-of-the-art autoscaling policies for workflows
are mostly plan-based or designed for batches (ensembles) of workflows.
This reduces their flexibility when dealing with workloads of workflows,
as the workloads are often subject to unpredictable resource demand fluctuations.
Moreover, autoscaling in clouds almost always imposes budget constraints
that should be satisfied. The budget-aware autoscalers for workflows usually require task
runtime estimates to be provided beforehand, which is not always possible when dealing
with workloads due to their dynamic nature.
To address these issues, we propose a novel Performance-Feedback Autoscaler (PFA) that is budget-aware and does not require task runtime estimates for its operation. Instead, it uses
the performance-feedback loop that monitors the average throughput on each resource type.
We implement PFA
in the popular Apache Airflow workflow management system, and compare the performance
of our autoscaler with other two state-of-the-art autoscalers, and with the optimal solution
obtained with the Mixed Integer Programming approach.
Our results show that PFA outperforms other considered online autoscalers, as it effectively minimizes the average job slowdown by up to 47\% while still satisfying the budget constraints.
Moreover, PFA shows by up to 76\% lower average runtime than the competitors.
\end{abstract}

\clearpage

\tableofcontents
\clearpage

\input{sections/introduction.tex}
\input{sections/problem-statement.tex}

\input{sections/autoscalers.tex}
\input{sections/experimental-setup.tex}
\input{sections/experimental-results.tex}
\input{sections/optimal-solution.tex}
\input{sections/threats-to-validity.tex}

\input{sections/related-work.tex}

\input{sections/conclusions.tex}

\appendix

\bibliographystyle{tudelft-report}
\bibliography{references}

\end{document}

%% file: sections/introduction.tex
\section{Introduction}\label{ch5:sec:introduction}
The variety of workflow structures observed in modern cloud workloads
and the diversity of cloud resource types require sophisticated
autoscaling policies for meeting Service Level Agreements (SLAs).
The problem of autoscaling for workflows previously has been often seen
just from the perspective of a single user who submits a batch (an ensemble)
of workflows to the cloud. Usually, it is supposed that the workflows in the batch have
previously known runtime characteristics, e.g., task runtime estimates, obtained through a code analysis,
a simulation, or by simply running the batch of workflows on a reference system.

This approach has been successfully adopted for executing batches
of scientific workflows~\cite{topcuoglu02, zhao06, malawski2015algorithms},
which are well-studied~\cite{juve2013characterizing} and have rather fixed patterns of execution~\cite{pegasuswfgenerator},
but can be too rigid for workflows in non-scientific domains~\cite{wen2017fog}.
Moreover, task runtime estimates have not been shown to be robust for batches in cloud settings, e.g., under multi-tenancy effects~\cite{jeyakumar2013eyeq} and performance variability~\cite{maricq2018taming}.

A more general approach assumes that the cloud user submits
a \emph{workload} of workflows of different types as, for example,
if the cloud user runs an application serving many other, diverse users of that application.
Many cloud-based services, such as Airbnb (rentals), Twitter (communication), and Netflix (video streaming), use this approach~\cite{oppitz2018cloud}.
In this situation, the service can be considered as a single cloud
user submitting a workload of workflow jobs.
Scheduling batches of workflows in the cloud normally has the goal to
minimize the makespan of the whole batch
and staying within the budget. However, in the workload
the number of arriving jobs can vary over time, thus, it is important
to minimize the workflow response time and slowdown, as both these
metrics include the possible queuing delay,
and look at the system from the stability perspective as
stability guarantees predictable and uninterrupted service.

Normally, a cloud user has some budget which expresses the financial
limitations for allocating the cloud resources. When dealing with
workloads of workflows, due to their dynamic nature, it is common that
the user budget is defined per certain time interval. Since cloud resources
are usually heterogeneous and have different costs, the basic goal
of the autoscaler in this case is to allocate a desired number of resources
and find such a combination of resource types that
maximizes the workload performance while staying within the budget constraint.
This also means that the resources can stay constantly running once allocated even
if they are idle. The deallocation of resources will only be needed
if that would help to adapt to the changing resource requirements of the workload, e.g.,
substitute lower number of expensive resources with higher number of cheaper ones
if that helps to increase the performance.
However, these requirements are not sufficient as multitenancy
in the cloud could lead to possible performance degradation from
a single user's perspective if there will be too many allocated but idle resources in the system.
Moreover, to address modern sustainability challenges and to minimize energy waste,
both end users and infrastructure owners should approach the use of resources responsibly.
For example, users could be interested in saving the unused budget
for achieving a sustainable business structure and minimizing the resource waste.
Similarly, the infrastructure owners could have the interest in providing
services that rely on sustainable infrastructure. 
In other words, it is not sufficient anymore to simply collect payments
for the resource usage without controlling how the resources are actually utilized.

Most of the existing autoscalers for workflows belong to the group of
\emph{offline} policies~\cite{abrishami2013deadline, malawski2015algorithms, arabnejad2017scheduling}.
Such policies, given a batch of workflows with known task runtimes,
create a full-ahead task placement and autoscaling plan which is then strictly followed
by the workflow management system that coordinates the execution.
The workflows could be submitted simultaneously or with some delays---the main distinguishing
feature of offline policies is that the submission times of all workflows are known in advance.
Moreover, in the work by Wang~et~al.~\cite{wang2017using} a Mixed Integer Programming (MIP)
approach was proposed that even allows to find the optimal solution to the autoscaling problem.

Only few papers consider the \emph{online} autoscaling scenario where
workflows arrive over time forming a workload~\cite{mao2011auto, byun2011cost, mao2013scaling},
and the arrival times are not known in advance.
However, such group of autoscalers mostly use the \emph{online plan-based} approach,
as they create a partial plan for the next autoscaling interval and not for the whole time horizon.
The main issue with various plan-based scheduling approaches is the excessive time complexity
they show when applied to workloads of workflows~\cite{versluis2018trace}.
That could negatively affect the stability of the system in case if the autoscaling decisions
take too much time, and do not scale well with workload fluctuations.
Shorter autoscaling intervals that are more in line with the current trend
on fine-grained billing~\cite{rodriguez2017budget} further complicate the problem, as plan-based autoscalers
simply do not have enough time for making their decisions.
Moreover, the plan-driven task placement can possibly delay the execution of
newly arrived workflows as the scheduler will need to wait until the new plan incorporating
the newly arrived tasks is constructed.
In this case, the plan-based computationally-intensive solutions are not beneficial
and can be substituted by simpler and faster heuristic approaches.

Thus, we clearly see further possibilities for improving autoscaling for workloads of workflows
by joining the concepts inherent both to the general~\cite{ali2012noms} and workflow-aware autoscalers~\cite{mao2013scaling}.
From general autoscaler we can use the performance-feedback mechanism and
the ability to derive and analyze runtime statistics during the execution.
For example, instead of trying to derive task runtime estimates~\cite{chirkin2017execution}, we can look at task throughput,
as the system can observe the task throughput fairly easily.
From workflow-aware autoscalers, instead of constructing a partial plan, we can use
less computationally intensive techniques for estimating the expected level of parallelism,
and, accordingly, the resource demand.

Thus, the main research questions in this paper are:
\begin{itemize}
    \item [Q1.] \emph{How to minimize workflow slowdowns within the budget constraint with unknown
    in advance task runtime estimates when autoscaling cloud resources for workloads of workflows?}
    \item [Q2.] \emph{Does the autoscaling policy found when answering Q1 has lower
    time complexity than the state-of-the-art plan-based online autoscalers?}
    \item [Q3.] \emph{How far is the performance and scalability of the policy
    found in Q1 from the optimal solution?}
\end{itemize}

When answering the raised questions, the contributions of this paper are the following:
\begin{enumerate}
    \item We answer Q1 by proposing a novel online dynamic Performance-Feedback Autoscaler (PFA)
    that uses the resource task throughput information and a token-based LoP estimator~(Section~\ref{ch5:sec:autoscalers}). 
    \item Through real-world experiments, PFA answers Q2 favourably
    by outperforming two state-of-the-art plan-based online autoscalers Planning First (PLF) and Scaling First (SCF)~(Section~\ref{ch5:sec:experimental-results}).
    \item We answer Q3 by comparing all the considered autoscalers
    with the optimal solution obtained from a Mixed Integer Programming model~(Section~\ref{ch5:sec:optimal-solution}).
\end{enumerate}

%% file: sections/problem-statement.tex
\section{Problem Statement}\label{ch5:sec:problem-statement}
This section presents the model for the problem of autoscaling for workloads of workflows.
The section also presents a set of metrics we use to evaluate the performance
of the workloads and the performance of the studied autoscalers.

\subsection{Autoscaling Model}\label{ch5:sec:model}
We consider a public cloud computing system which is
a subject to an arriving workload of workflows.
The workload consists of multiple independent sub-workloads
each belonging to an independent user.
Each job in the workload is a workflow, and each
component of the workflow is a task.
Each workflow has single entry task and single exit task.
The tasks belonging to a single workflow can exchange data
through a shared file system.
The task is considered \emph{eligible} when all of its precedence constraints are satisfied,
e.g., when all of its required input files are available in the shared file storage.
The \emph{workflow size} is the total number of tasks in a workflow.

The cloud computing system allows every user to dynamically
allocate and deallocate computing resources of various
types, where each resource type has a specific cost.
Each resource can be in either of the following four states:
\emph{down}, \emph{idle}, \emph{busy}, or \emph{booting}.
The resource is \emph{down} when it is deallocated and it is not reserved for any user.
The resource is \emph{idle} when it is allocated,
reserved for a certain user, but has no currently assigned task.
The resource is \emph{busy} when it has a task assigned.
The resource is \emph{booting} when it is in the transition state between the down and idle states.
Once the resource is allocated, the user is charged and the resource
is reserved for the user until the end of the resource billing period, where
the \emph{billing period} is the minimal time for which
the cloud resource can be reserved for a particular user.
Each user has a certain operational budget per autoscaling interval,
and the total cost of all the resources reserved for the user
on the autoscaling interval cannot exceed the user budget.
After the allocation, the resource spends some time in the booting state,
while already being reserved for the user, without being able to execute any user tasks.
At the end of the billing period, the resource can be deallocated or its
reservation can be prolonged for the next billing period.
In our model, the duration of the billing period equals to
the duration of the autoscaling period.
Before transitioning into the down state for deallocation,
the resource should always pass the idle state first.
The resource deallocation happens instantaneously.
The \emph{system size} is the maximal resource capacity
which is available for the system users.

Since the number of eligible tasks from each user varies over time,
the system employs an \emph{autoscaler} to automatically control the number
of allocated resources on per-user basis.
The separate \emph{scheduler} is responsible for placing tasks onto the allocated resources.
In this work, we focus on periodic autoscaling, so that the autoscaler
is invoked at fixed intervals by the workflow management system
and monitors the controlled cloud environment.
Accordingly, the \emph{autoscaling interval} is the time between
any two invocations of the autoscaler.

Despite that the system has resources of different types,
we assume that there is no direct dependency between the cost of a resource type and
the execution speeds of tasks running on it. Our motivation is based on the assumption
that while some tasks can benefit from additional CPU cores, other tasks
can be sequential in their nature and, thus, can show better performance
on resources with fewer cores but with higher CPU frequency.
Similar assumptions can be made for different RAM or storage requirements, etc.

The autoscaler and the scheduler operate in tandem with the goal to
\emph{minimize workflow response time within the budget constraint}.
This can be achieved by allocating enough resources and finding an appropriate
resource profile which guarantees required performance.
By \emph{resource profile} we understand a specific combination
of resource types within the set of resources currently allocated for the user.
Additionally, the autoscaler can have a goal to achieve fairness among multiple users.
Since the resources are reserved for each user until the end of their billing period,
during that period each resource can execute only tasks from the reserving user.
This implies that the scheduler is not able to control the fairness among
the users as it is only allowed to place tasks belonging to a certain user to the
resources that are reserved for the same user. Thus, the only way to control fairness,
by which in this chapter we understand maintaining average task throughput
proportional to the user budget, is by controlling the number
of allocated resources within the resource profile.
Thus, if the autoscaler is fairness-aware, it should consider
in addition to the budget constraint also the fairness constraint.

We do not include deadlines in this study as, in contrast to the offline approach,
in the dynamic workload scheduling deadlines can only be roughly estimated.
Furthermore, for workloads the deadline compliance depends on the system utilization, thus,
the deadlines that were derived previously at a certain utilization level
can be easily invalid for other utilizations.
The dynamic nature of autoscaling for workloads makes the response time minimization and
the stability of the system more important goals rather than the deadline compliance.
It is also reasonable to assume that the response time minimization usually increases
the number of met deadlines.
Additionally, in our model we allow users to assign
numeric priorities to workflows so that they can indicate which workflows
are more important and should be processed faster.

\subsection{Performance Metrics for Autoscaling}\label{ch5:ssec:performance-metrics}
The system is constantly monitored by its users and operators, who assess its performance for a set of metrics commonly used in autoscaling settings~\cite{ilyushkin2018experimental}.

\subsubsection{User- and System-Oriented Metrics}\label{ch5:sssec:performance-metrics}
As a main user-oriented metric we use \emph{slowdown} 
which is defined in steps as follows:
The \emph{waiting time} is the time that a workflow spends in the system before
starting executing its first task.
The \emph{makespan} is the time between the start of
the first task of the workflow and until the completion of its last task.
The \emph{response time} of a workflow is the sum of its waiting time and its makespan.
The \emph{slowdown} is the ratio of response time of a workflow in a busy system
to the ideal workflow makespan obtained from a reference system.
We also consider the \emph{monetary cost per autoscaling interval}
as a user-oriented metric. By monetary cost we understand
the total cost of the allocated resources during any autoscaling interval.
As a system-oriented metrics we use the \emph{percentage of busy resources}
throughout the experiment and the \emph{percentage of allocated resources
per autoscaling interval}.

\subsubsection{Elasticity-Oriented Metrics}\label{ch5:sssec:elasticity-metrics}
To evaluate the performance of the considered autoscaler, we take the elasticity into account. In our model, we allow each resource to run only a single workflow task at a time. Accordingly, the momentary \emph{demand} equals to the number of currently running and eligible workflow tasks (submitted by a particular user). By the resource \emph{demand} $d_t$ we understand the minimal number of resources required for fulfilling a given performance-related Service Level Objective (SLO) at time \mbox{$t \in [1,T]$}.
By the resource \emph{supply} $s_t$ we understand the number of currently allocated (to the user) resources which are not in the down state at time $t \in [1,T]$.
The maximal number of resources that can be supplied $R$ is limited.

The \emph{under-provisioning accuracy} $a_U$ is defined as the average fraction of missing resources required to meet the SLO. Similarly, the \emph{over-provisioning accuracy} $a_O$ is the average fraction of resources that the autoscaler supplies in excess of the current demand. Both metrics can be formulated as: 
\begin{align}
    &a_{U} = \frac{1}{T \cdot R} \cdot \sum_{t=1}^T\max(d_t - s_t,0),   \\
    &a_{O} = \frac{1}{T \cdot R} \cdot \sum_{t=1}^T\max(s_t - d_t,0).  
\end{align}

The \emph{under-pro\-visioning time share} $t_U$ is the time relative to the measurement duration, in which the system has insufficient resources, whereas, the \emph{over-provisioning time share} $t_O$ is the time relative to the measurement duration, in which the system has more resources than required. Both metrics can be computed as:
\begin{align}
    &t_{U} = \frac{1}{T} \cdot \sum_{t=1}^T\max(\sgn(d_t - s_t),0), \\
    &t_{O} = \frac{1}{T} \cdot \sum_{t=1}^T\max(\sgn(s_t - d_t),0).
\end{align}

%% file: sections/autoscalers.tex
\section{Autoscalers}\label{ch5:sec:autoscalers}
This section explains in detail two state-of-the-art budget-aware autoscalers, 
that require task runtime estimates for their operation, and
presents our novel autoscaler, which, in contrast,
operates without explicitly provided task runtime estimates.
The considered state-of-the-art autoscalers were proposed by Mao and Humphrey~\cite{mao2013scaling}
and designed specifically for workloads of workflows.
The relevance of these autoscalers is supported by the recent survey~\cite{lu2018review}.

\subsection{Planning-First Autoscaler}\label{ch5:ssec:scheduling-first-autoscaler}
The Planning First (PLF)~\cite{mao2013scaling} autoscaling policy uses currently eligible tasks to allocate resources within a budget constraint.
Even though the name of the policy in the original paper is Scheduling First,
further we refer to it as Planning~First,
as this policy basically creates an execution plan for the tasks within the autoscaling interval.
The autoscaler consists of six steps which are executed on every policy invocation, i.e., for every autoscaling interval:
\begin{enumerate}
    \item[i.] Distribute the user budget among the workflows based on their priority.
    \item[ii.] Perform initial supply prediction by determining the number of each resource type to allocate within the budget constraint.
    \item[iii.] Consolidate the budget left after the initial supply prediction.
    \item[iv.] Allocate the resources according to the predicted supply.
    \item[v.] Create an execution plan for the upcoming autoscaling interval.
    \item[vi.] Deallocate idle resources which do not have any tasks planned and are approaching the end of their billing period.
\end{enumerate}

In the first step, the policy computes the cost of already allocated resources,
deducts their cost from the user budget, and distributes the remaining budget to individual workflows
proportionally to their priority, so that higher priority workflows get bigger budgets.

In the second step, the policy iterates through the eligible tasks of the workflow,
sorted in the descending order of their workflow priorities, and for each task, while there is enough budget,
it finds the resource type allowing to finish the task in the shortest time.
The tasks are not assigned to the resources, only the number of resources of each type is determined.
If the budget is over, the autoscaler proceeds to the third step---the budget consolidation.
In the original paper, the loop break condition depends on the cost of the cheapest resource in the system
so that already after the second step the policy can overspend the budget (for each workflow)
by the cost difference between the fastest resources and the cheapest one.
To avoid this, we modify the policy and use the cost of the fastest resource
for the currently processed workflow task instead.

In the third step, the policy performs budget consolidation, as some budget can be left
by individual workflows after the initial supply prediction. There are two reasons why
the initially distributed budget may not be fully spent:
some workflows could have not enough eligible tasks, or
some workflows could have remaining budget smaller than the cost of the fastest resource.
So that these remaining per-workflow budgets can be redistributed among the workflows from the same user
to include more fastest resources in the allocation plan.
This allows to determine fastest resource types
for the remaining higher priority eligible tasks that were not processed in the second step.
After this step, the autoscaler produces the final predicted number of instances of each type which should
be allocated. It also specifies for some or all eligible tasks on which resource types they
should run. Some eligible tasks belonging to lower priority workflows still could be
without assigned resource types, as the cost of their fastest resources
did not fit within the budget constraint.

In the fourth step, the policy performs so-called resource consolidation which basically means creation
of an execution plan on the already allocated (at the moment of the autoscaler invocation)
and newly allocated resources (after the third step) for the upcoming autoscaling interval.
For that, the policy determines actual resources (not just the resource types) for each workflow task
and tries to fill the resources in the plan with tasks until the end of the autoscaling interval.
This is necessary, as after the third step only (a subset of) eligible tasks
get the resource type assigned---those, that were used to predict the supply.
Accordingly, the number of resources in the plan equals the number of running tasks and the number
of tasks that have the resource type assigned after the third step.
As the original paper used simulations, many very important details, which are crucial
when implementing the policy in a real system, are missing or imprecise.
Further we provide our interpretation of the resource consolidation step.

In the fourth step, the policy allocates the resources according to the predicted supply.

In the fifth step, the policy performs resource consolidation, i.e., it creates
a task placement plan for the upcoming autoscaling interval, while processing the workflows in the random order.
The newly allocated resources are considered as booting, thus, the planner takes into account
the allocation delay which is supposed to be known in advance.
The execution plan is initialized with tasks that are already running at the moment of the autoscaler invocation.
Then the policy adds in the plan the eligible tasks that got the resource type assigned during
the second or third autoscaling steps.
The eligible tasks with known resource types are first assigned to idle resources of that type.
If there are no idle resources, the planner checks the booting and busy resources of the same type,
which of those will become available earlier, and places the eligible tasks on the earliest one.
After that, all the resources in the plan should have at least one task assigned.
Finally, all the remaining eligible and not yet eligible tasks are processed
while maintaining the precedence constraints, i.e., a task is added to the plan if all of its parents are already in the plan.
Each task is placed on the resource which is at the moment of task placement provides the minimal earliest possible start time.
The planning process continues until there are no tasks that
can start their execution before the end of the autoscaling interval.

Finally, in the sixth step, the resources that did not get any tasks assigned in the previous steps and that
are approaching the end of their billing period are deallocated.

\subsection{Scaling-First Autoscaler}\label{ch5:ssec:scaling-first-autoscaler}
The Scaling First (SCF)~\cite{mao2013scaling} autoscaling policy first creates for each workflow an individual
execution plan (without considering resource allocation constraints),
and then scales the plan so that it fits within the user budget constraint.
The policy consists of five major steps:
\begin{enumerate}
    \item[i.] Perform initial supply prediction by creating a per-workflow execution plan without limiting the number of resources.
    \item[ii.] Scale the initial prediction to fit within the budget constraint, and consolidate the remaining budget.
    \item[iii.] Allocate the resources according to the predicted supply.
    \item[iv.] Create an execution plan for the upcoming autoscaling interval.
    \item[v.] Deallocate idle resources in the same way as in the PLF policy.
\end{enumerate}

In the first step, the policy creates an independent (from other workflows) per-workflow plan neither considering
the system resource allocation limits nor considering the budget constraint.
Thus, the number of resources in each plan can be bigger than the actual number of maximally available resources in the system.
Since the original paper does not clearly explains this step, we present our detailed interpretation
of the procedure for creating the per-workflow plan which uses similar logic as the resource consolidation step.
First, the policy selects all the already running tasks of the current workflow
and places all of them in the plan. Their resource types are already known, as well as the expected finish times.
Second, the policy selects all the eligible tasks and places them on their fastest
resource types, calculating the appropriate expected finish time.
Third, all the other not yet eligible tasks are placed in the plan on their fastest resources
(if those required fastest resources are not yet in the plan then they are added)
so that the earliest possible start time for each task is minimized at the moment of its addition to the plan.
Similarly to the resource consolidation step of PLF, a task is added to the plan only if all of its parents are already in the plan.
The final number of resources for each resource type that should be supplied is calculated as the rounded up sum of
the runtimes of planned tasks on each resource type divided by the length of the autoscaling interval.

In the second step, for each resource type the policy proportionally scales the initially predicted supply
by multiplying it by the factor calculated as the fraction of the user budget and the total cost of initially predicted resources.
Since the number of resources is integer, some remaining budget can be left after scaling the initial supply.
This remaining budget is used to allocate more resources, if possible.
For that the policy iterates in a round robin manner through the predicted in the first step resource
types until even the resource of the cheapest type cannot be allocated.

In the third step, the policy allocates the resources according to the predicted supply.

In the fourth step, the policy performs resource consolidation.
We modify the resource consolidation approach described in the original paper for SCF, as it is does not mention
the situation when the number of resources of a certain type after the scaling step is zero.
Instead, we use the approach similar to our interpretation of resource consolidation for the PLF policy.
There are two differences between SCF and PLF. First, in PLF before the resource consolidation step
some (or all) eligible tasks have the resource type already assigned, while in SCF
the information on the preferred resource types from the first step is completely discarded.
Second, in SCF the tasks are added to the plan in the order of their workflow priorities,
so that higher priority tasks are added to the plan earlier.

The fifth step of the SCF policy is identical to the resource deallocation step of PLF.

\subsection{Performance-Feedback Autoscaler}\label{ch5:ssec:performance-feedback-autoscaler}
In this section, we present our novel Performance-Feedback Autoscaler (PFA), which we developed
considering the limitations of the state-of-the-art workflow-specific autoscalers, and
based on observations on the performance of general and workflow-specific autoscalers from the literature~\cite{malawski2015algorithms, ilyushkin2018experimental}.

We expect PFA to achieve better elasticity performance, as it
constantly monitors the historical resource throughputs
to derive faster resource types,
and relies on a low complexity workload approximator
to predict the future demand.
Moreover, the dynamic task placement, used together with PFA,
is expected to further reduce task waiting times
and increase the resource utilization.

The PFA autoscaler consists of the following steps:
\begin{enumerate}
    \item[S1.] Determine the resource profile using the historical throughputs.
    \item[S2.] Determine the number of resources (the supply) that can be allocated within the resource profile with the user budget.
    \item[S3.] Estimate future workload resource demand using the token propagation approach and historical throughput information.
    \item[S4.] Scale down the profile-based supply if it is higher than the predicted demand to avoid wasting resources.
    \item[S5.] Allocate the predicted number of resources.
    \item[S6.] Deallocate idle resources which are staying idle the longest and approaching the end of their billing period.
\end{enumerate}

\begin{table}[!ht]
\centering
\caption{Symbols used for the PFA autoscaler.}\label{ch5:tab:pfa-symbols}
\begin{tabularx}{\linewidth}{lX}
\toprule
\multicolumn{2}{c}{\textbf{Inputs}}\\
\hline
$t$ & The autoscaling interval, $t \in \mathbb{Z}_{\geq 0}$, where $t = 0$ corresponds to the earliest autoscaling interval\\
$m$ & The lookup depth for MA and TBA, $m \in [0, t]$\\
$\alpha$ & The EWMA smoothing factor, $\alpha \in [0, 1)$\\
$\mathcal{R}$ & The set of resource types, $i \in \mathcal{R}$\\
$\mathcal{U}$ & The set of users, $j \in \mathcal{U}$\\
$q_i$ & The resource cost on any single autoscaling interval\\
$b_j$ & The user budget for a single autoscaling interval\\
\hline
\multicolumn{2}{c}{\textbf{System Measurables}}\\
\hline
$\tau_{i,j}(t)$ & The average throughput\\
$c_{i,j}(t)$ & The number of completed tasks\\
$n_{i,j}(t)$ & The number of allocated resources\\
\hline
\multicolumn{2}{c}{\textbf{Derived Values}}\\
\hline
$\hat{\rho}_{i,j}(t)$ & The instant resource type ratio\\
$\rho_{i,j}(t)$ & The smoothed resource type ratio\\
$\nu_{i,j}(t)$ & The budget fraction available for the resource type\\
$\zeta_j(t)$ & The lookup depth for the token-based approximator\\
$\theta_j(t)$ & The number of tasks in the visited future eligible sets\\
$\lambda_j(t)$ & The token-approximated LoP\\
$\sigma_j(t)$ & The token-approximated demand for all resource types\\
$\hat{\mu}_{i,j}(t)$ & The throughput-based number of resources to allocate\\
$\tilde{\mu}_j(t)$ & The throughput-based number of resources to allocate\\
$\mu_{i,j}(t)$ & The final corrected number of resources to allocate\\
$P_{i,j}(t)$ & The history of non-zero total resource ratios for MA\\
$T_j(t)$ & The history of non-zero total throughputs for MA used with TBA\\
\bottomrule
\end{tabularx}
\end{table}

\subsubsection{Determining the Resource Profile}
\label{ch5:sssec:pfa-resource-profile}

The first two steps of the autoscaler use throughput information to derive
the initial resource profile. PFA relies on two alternative smoothing mechanisms
for the historical throughput: Moving Average (MA) and Exponentially Weighted Moving Average (EWMA) for smoothing out possible throughput fluctuations.

In the first step, on the autoscaling interval $t$ for each resource type $i$
and each user $j$ the average resource throughput $\tau_{i,j}(t)$ is defined as:
\begin{equation}
    \tau_{i,j}(t) =
    \begin{cases}
    \frac{ c_{i,j}(t) }{ n_{i,j}(t) }, & \text{if } n_{i,j}(t) > 0, \\
    0, & \text{otherwise,}
    \end{cases}
\end{equation}
where $c_{i,j}(t)$ is the number of completed tasks on the interval, 
and $n_{i,j}(t)$ is the number of allocated resources on the interval.
This allows to compute the instant throughput-based resource type ratios:
\begin{equation}
  \hat{\rho}_{i,j}(t) =\begin{cases}
    \dfrac{\tau_{i,j}(t) }{ \sum\limits_{r \in \mathcal{R}} \tau_{r,j}(t) }, & \text{if } \sum\limits_{r \in \mathcal{R}} \tau_{r,j}(t) > 0,\\
    0, & \text{otherwise,}
  \end{cases}
\end{equation}
where $\mathcal{R}$ is the set of all the resource types in the system.
MA uses resource type ratios that are not zero for all the resource
types:
\begin{equation}
    P_{i,j}(t) = \bigg\{ \hat{\rho}_{i,j}(t-k) : \sum\limits_{r \in \mathcal{R}} \hat{\rho}_{r,j}(t-k) > 0, \forall k \in [0, m] \bigg\}.
\end{equation}
The MA-smoothed resource ratios
over $m$ previous observations are computed as:
\begin{equation}
    \rho_{i,j}(t) =\begin{cases}
    \frac{1}{|P_{i,j}(t)|}\cdot\sum\limits_{k \in P_{i,j}(t)}^{} k, & \text{if } \sum\limits_{k \in P_{i,j}(t)}^{} k > 0,\\
    \frac{1}{|\mathcal{R}|}, & \text{otherwise},
  \end{cases}
\end{equation}
where $|X|$ denotes the cardinality of a set $X$. If any resource has zero historical throughput
all the resource types, instead, get an equal share. 
This allows the system to collect
the throughput history for all the resource types.
For the EWMA smoothing method, the smoothed resource ratios are computed as:
\begin{equation}
    \rho_{i,j}(t) =\begin{cases}
    \alpha \cdot \rho_{i,j}(t-1) + (1-\alpha) \cdot \hat{\rho}_{i,j}(t), & \text{if } \hat{\rho}_{r,j}(t) > 0, \forall r \in \mathcal{R}, \\
    \frac{1}{|\mathcal{R}|}, & \text{otherwise},
  \end{cases}
\end{equation}
with $\alpha \in [0,1)$ being the smoothing factor. The parameter $\alpha$ represents the degree of weighting decrease of the past values of $\rho_{i,j}$. A small value of $\alpha$ (close to $0$) corresponds to the non-averaged value of $\rho_{i,j}$, i.e., $\hat{\rho}_{i,j}$, while a high value (close to $1$), corresponds to a smoother signal over time.

In the second step, based on the resource ratio produced in the first step,
we calculate the number of resources of each type that can be allocated with the user budget. For that, we define the fraction $\nu_{i,j}(t)$ of the user budget
that we can spend on each resource type according to the resource ratio, knowing the cost $q_i$ of each resource type $i$: 
\begin{equation}
    \nu_{i,j}(t) = \frac{ q_i \cdot \rho_{i,j}(t) }{ \sum\limits_{r \in \mathcal{R}} \big( q_r \cdot \rho_{r,j}(t) \big) }.
\end{equation}
Accordingly, for each user $j$ the number of resources of type $i$
that can be allocated with the user budget $b_j$ is calculated as:
\begin{equation}
    \hat{\mu}_{i,j}(t) = \left\lfloor \frac{b_j \cdot \nu_{i,j}(t)} {q_i} \right\rfloor,
\end{equation}
supposing that the budget is large enough to allocate at least one instance of each resource type, where $\left\lfloor x \right\rfloor$ denotes the floor function of a real number $x$.
Summing up the $\hat{\mu}_{i,j}(t)$ values for all the resource types we calculate the total resource supply that can be achieved with the obtained resource profile:
\begin{equation}
\tilde{\mu}_j(t) = \sum_{r \in \mathcal{R}} \hat{\mu}_{r,j}(t).
\end{equation}

\subsubsection{Token-based Demand Prediction}
\label{ch5:sssec:pfa-token-based-prediction}
In the third step, the resource demand $\sigma_j(t)$ is predicted
using the Token-based Approximator (TBA), similar to the one described in~\cite{ilyushkin2018experimental}. 
For that, TBA considers all the submitted and not yet
finished workflows of the user as
a single workflow, excluding finished tasks, and places tokens in all the tasks that
either have no parents or whose parents has already finished.
Then in successive steps TBA moves these tokens to all the tasks all of whose parents
already hold a token or were earlier tokenized. TBA records the total number
of token movements and, after each step, the number of tokenized nodes.

The intuition is to evaluate the number of ``waves''
of tasks (future eligible sets)
that will finish during the autoscaling interval.
When the lookup depth $\zeta_j(t)$ or
the final task of the joint workflow is reached,
the largest recorded number of tokenized nodes is
the approximated LoP $\lambda_j(t)$, and the total number of token movements
$\theta_j(t)$ is the number of tasks in the visited future eligible sets.
To limit the TBA lookup depth $\zeta_j(t)$, we use the average historical task throughput
among all the resource types smoothed either with MA over $m$ previous autoscaling intervals, or with EWMA.
For MA, the set of historical average throughputs for all the resource types
with skipped intervals with zero total throughput is defined as:
\begin{equation}
    T_{j}(t) = \bigg\{\tau_{i,j}(t-k) : \sum\limits_{r \in \mathcal{R}} \tau_{r,j}(t-k) > 0,
    \forall k \in [0, m], \forall i \in \mathcal{R} \bigg\},
\end{equation}
With MA, the TBA lookup depth is calculated as:
\begin{equation}
    \zeta_j(t) =
    \begin{cases}
    \Big\lceil \frac{1}{|T_{j}(t)|} \cdot \sum\limits_{k \in T_{j}(t)}^{} k \Big\rceil, & \text{if } \sum\limits_{k \in T_{j}(t)}^{} k > 0,\\
    \infty, & \text{otherwise,}
  \end{cases}
\end{equation}
where $\left\lceil x \right\rceil$ denotes the ceiling function of a real number $x$.
Accordingly, the resource demand $\sigma_j(t)$ with MA is computed as:
\begin{equation}
    \sigma_j(t) = 
    \begin{cases}
    \left\lceil \theta_j(t) \cdot |T_{j}(t)| \cdot \left(\sum\nolimits_{k \in T_{j}(t)}{} k\right)^{-1} \right\rceil, & \text{if } \sum\limits_{k \in T_{j}(t)}^{} k > 0,\\
    \lambda_j(t), & \text{otherwise.}
    \end{cases}
\end{equation}
With EWMA, the TBA lookup depth is calculated as:
\begin{equation}
    \zeta_j(t) =
    \begin{cases}
    \bigg\lceil \alpha \cdot \zeta_j(t-1) + (1 - \alpha) \cdot \frac{\sum\limits_{r \in \mathcal{R} }^{}\tau_{r,j}(t)}{|\mathcal{R}|} \bigg\rceil, & \text{if } \sum\limits_{r \in \mathcal{R} }^{}\tau_{r,j}(t) > 0,\\
    \infty, & \text{otherwise.}
  \end{cases}
\end{equation}
And the resource demand $\sigma_j(t)$ with EWMA is computed as:
\begin{equation}
    \sigma_j(t) = 
    \begin{cases}
    \Big\lceil \theta_j(t) \cdot |\mathcal{R}| \cdot \Big(\sum\limits_{r \in \mathcal{R} }^{}\tau_{r,j}(t)\Big)^{-1} \Big\rceil, & \text{if } \sum\limits_{r \in \mathcal{R} }^{}\tau_{r,j}(t) > 0,\\
    \lambda_j(t), & \text{otherwise.}
    \end{cases}
\end{equation}

\subsubsection{Scaling Down or Inflating the Profile}
\label{ch5:sssec:pfa-profile-scaling-down-inflating}
In the fourth step, we scale down or inflate, if necessary,
the resource supply calculated using the resource profile
to match the predicted resource demand $\sigma_j(t)$.
Scaling down prevents allocation of potentially idle resources,
and gives space to other users to utilize the resources.
Inflating the profile, despite creating
possible imbalance in the throughput-based resource ratio,
helps to cope with sudden demand surges
by increasing the total throughput.
If $\tilde{\mu}_j(t)$ exceeds the predicted demand $\sigma_j(t)$,
we proportionally scale down the $\hat{\mu}_{i,j}(t)$ values:
\begin{equation}
    \mu_{i,j}(t) = \left\lceil \frac{\sigma_j(t)}{\tilde{\mu}_j(t)} \cdot \hat{\mu}_{i,j}(t) \right\rceil.
\end{equation}
If $\tilde{\mu}_j(t)$ is lower than the predicted demand $\sigma_j(t)$,
we inflate the resource as follows:
{(i)}~Sort the resources in the ascending order
of their resource type cost.
{(ii)}~For each resource type $i$, except the most expensive one,
try to add to the original resource profile $\hat{\mu}_{i,j}(t)$ as many
resources of that type as possible, until there is no budget available or until $\tilde{\mu}_j(t)$ reaches $\sigma_j(t)$.
This produces the inflated $\mu_{i,j}(t)$ values.
{(iii)}~If $\sigma_j(t)$ is not yet reached,
starting from the second cheapest resource $k$,
try to remove one instance of it from $\mu_{k,j}(t)$
and, instead, add a number of instances to the previous cheapest
resource type $\mu_{k-1,j}(t)$. This does not change the total
cost of the resource profile, but increases $\tilde{\mu}_j(t)$.
Continue, until the total number of resources
in the profile reaches $\sigma_j(t)$ or no more such exchanges are possible.

\subsubsection{Allocating and Deallocating Resources}
\label{ch5:sssec:pfa-resource-allocation-deallocation}
In the fifth step, the predicted number of resources is allocated according
to the $\mu_{i,j}(t)$ values while taking into account the already allocated
resources and the physical system constraints.
In the sixth step, PFA de-allocates at maximum the number of idle resources that exceeds the predicted supply.
While de-allocating the idle resources, PFA gives priority to those that approach the end of their billing interval.

\subsubsection{Task Placement}
\label{ch5:sssec:pfa-task-placement}
The PFA autoscaler can operate with various independent task placement policies.
As both of the state-of-the-art autoscalers considered in this work, PLF and SCF, employ user-defined workflow priorities and both
have embedded task-placement policies which construct an execution plan,
for comparability purposes together with PFA we use dynamic task placement policy which
also considers user-defined workflow priorities. Our task placement policy
assigns eligible tasks according to the priority of their workflows
to the first available idle resource of any type.

%% file: sections/experimental-setup.tex
\section{Experiment Setup}\label{ch5:sec:experimental-setup}
This section describes the setup we used to conduct the experiments
and the synthetic workloads of workflows.
 
\subsection{Apache Airflow Deployment and Configuration}\label{ch5:ssec:airflow-deployment-configuration}

\begin{figure}[!t]
\centering
    \includegraphics[width=0.8\linewidth]{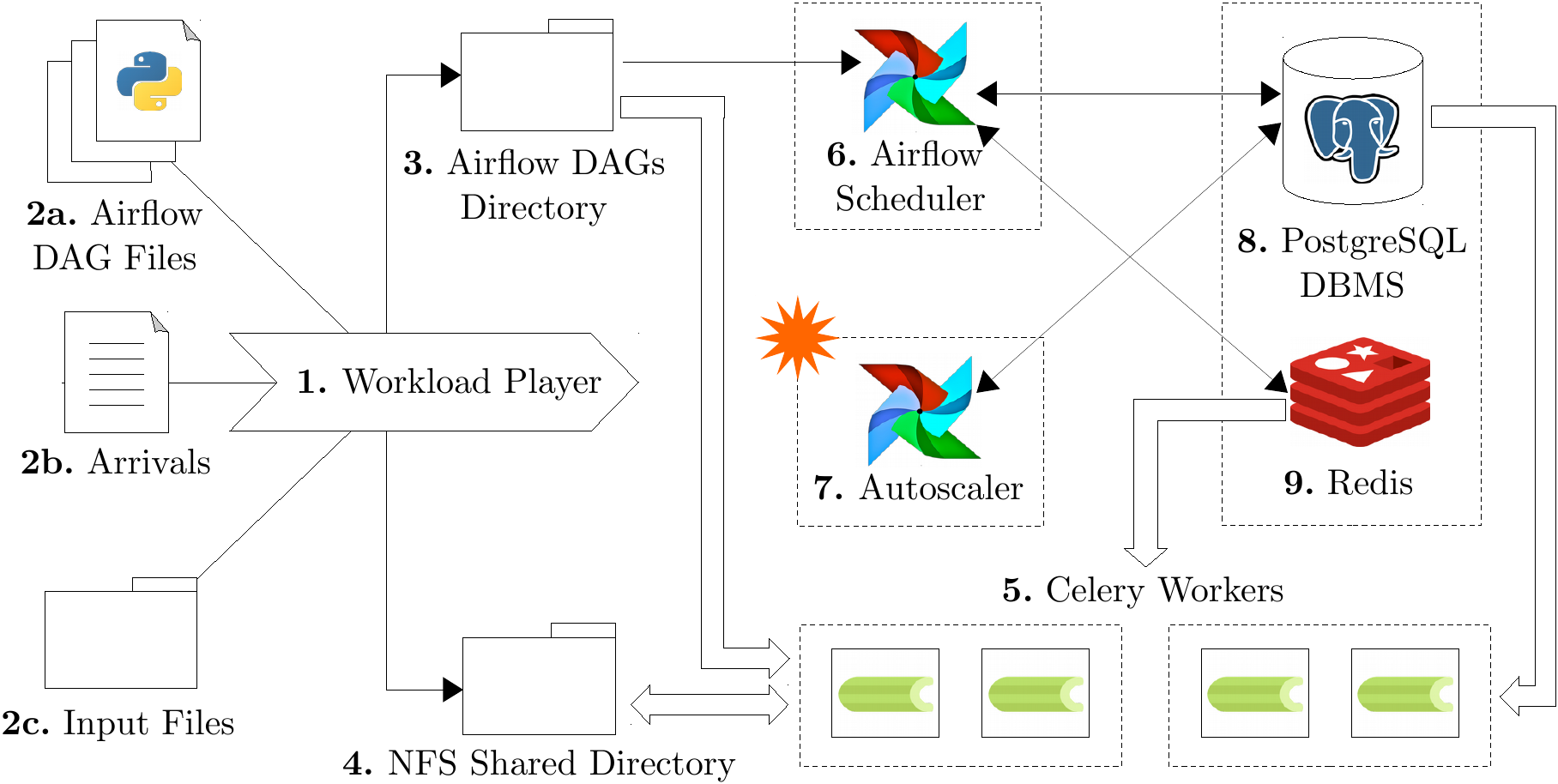}
    \caption{The architecture of the system.
    \label{ch5:fig:system-architecture}}
\end{figure}

Our setup is based on the Apache Airflow WMS~\cite{airflow}
(v1.9.0) which we extended
by adding an autoscaling component with a resource manager.
We choose Airflow since it is open source, it is written in Python, and it uses Python-based workflow descriptors, making it rather easy to integrate our code using the existing Airflow codebase.
Airflow has reasonable performance for running workloads of workflows and for the autoscaler evaluation purposes.
Moreover, Google provides Airflow as its Cloud~Composer service~\cite{googlecomposer}.
The architecture of our system is presented in Figure~\ref{ch5:fig:system-architecture}.
All the components of the system are deployed on a cluster with the following characteristics.
Head node: Intel Xeon X5650~@~2.67GHz CPU, 49GB RAM, 18TB HDD.
32 compute nodes: Intel Xeon E5620~@~2.40GHz CPU, 24GB RAM, 2TB HDD.
The cluster employs the QDR InfiniBand interconnect and 1 Gbit/s Ethernet at the compute nodes and 10 Gbit/s Ethernet on the head node. All the nodes are running \mbox{CentOS} (v7.4.1708).
The average measured Network File System (NFS) access speed is 550 MB/s.

The Workload~Player (Component~1 in Figure~\ref{ch5:fig:system-architecture})
emulates the Poisson workflow arrivals
by sequentially copying workflow descriptors (Component~2a) to the Airflow DAGs directory (Component~3)
according to the interarrival times which are read from the Arrivals file (Component~2b).
The interarrival times are pre-generated knowing the average
total workflow execution time in the workload and the size of the system,
so that the imposed average system utilization is kept around 20\%.
We choose this relatively low imposed utilization to
better evaluate the considered autoscalers as it minimizes
the amount of time when the demand significantly exceeds
the maximal achievable supply.
When the descriptor appears in the Airflow DAG directory, the Workload~Player issues the \texttt{`trigger\_dag'} Airflow command
to start the workflow execution.
In each workflow descriptor we define an identifier of the user who owns the workflow. 
Together with the workflow descriptor, the Workload~Player copies
the required input files (Component~2c) to a shared directory in the Network File System (NFS) (Component~4)
which is accessible to all the cluster nodes. The Airflow system does not provide
specific interface for accessing workflow files, thus, the workflow code
is responsible for file access operations.
Each task can start its execution when all of its input files are read.
Similarly, each task is considered as finished when all its output files are written.
The minimal delay between any two dependent tasks is equal to the sum of these two values.

All the Airflow components communicate through the central
Airflow database which is in our setup deployed in the PostgreSQL database
management system (Component 8).
Our setup uses the Celery~\cite{celery} distributed task queue (version 4.1.1)
with Redis~\cite{redis} (Component~9) in-memory database (version 4.0.10)
for sending tasks to the worker nodes.
Each worker node runs 8 Celery workers (Component~5)---one per CPU core.
In total we deploy 64 Celery workers on 8 worker nodes.

The Airflow~Scheduler (Component~6) is responsible for placing eligible tasks
for execution to the resources (Celery workers).
The default Airflow scheduler is an online dynamic scheduler
as it simply sends the eligible tasks in the order of their priority
to the single Celery queue which is monitored by the worker processes.
Even though, Airflow supports pools of workers, it does not have functionality
to monitor the status of each individual worker,
and does not support assigning workers to users.
To implement this functionality we introduce individual Celery queues
for each worker (resource) and guarantee that no new task is placed in the resource's queue
if the queue is not empty, so that the queue can hold one task at maximum.
We add a table in the central Airflow database 
which, for each queue (i.e., resource), stores the information on its current status
and the identifier of the user who reserved the resource.
Such an approach is required since PLF and SCF autoscalers are not only
responsible for the resource allocation but also partially take
the work from the scheduler by constructing the tasks placement plan
for the whole autoscaling interval. Thus, for PLF and SCF autoscalers
our Airflow~Scheduler simply places tasks to the idle resources
just according to the plan. However, since our PFA autoscaler does not create any plan,
the modified Airflow~Scheduler, when working in tandem with PFA,
makes its own task placement decisions by sending
eligible tasks according to their workflow priority and task priority
to the first idle resource.
In all the cases, the modified Airflow~Scheduler only places tasks that belong
to a specific user to the resources that are reserved for the same user. 

The Autoscaler (Component 7) 
is a novel independent component
which is implemented from scratch but heavily relies on the existing Airflow codebase.
The Autoscaler implements all the three considered autoscaling policies which
can be configured through the main Airflow configuration file. The Autoscaler
monitors the status of resources and changes their status through the central Airflow database.

The Workload~Player, Airflow~Scheduler, and Autoscaler are running on individual worker nodes. The PostgreSQL database management system is co-located
with the Redis in-memory database on the the head node.

\subsection{Billing Setup}\label{ch5:ssec:billing-setup}
We configure the system with two Small and Large resource types with different costs.
Further, we use the generic currency sign \textcurrency~when referring to monetary costs. 
An instance of Small resource type costs 1\textcurrency~per billing interval,
while a Large instance costs 5\textcurrency~per billing interval.
The system is configured to allocate at maximum 64 resources of which 32 of type Small and 32 of type Large.
Accordingly, the maximal budget that all the users can spend per autoscaling
interval is 192\textcurrency.
If both users have joint budget which allows to purchase more resources than the system can provide
the users will be competing between each other.
We shuffle the users before executing the autoscalers for each of them.
In the simple case when the system would have only a single resource type
any of the considered autoscalers would not make sense as each user would simply get
the number of instances which its budget allows to allocate at maximum.
We use autoscaling interval of one minute to be in line
with the current trend on fine-grained billing~\cite{rodriguez2017budget}.
Since we report the imposed system utilization, we believe that the same behavior should be observed
for shorter or longer autoscaling intervals, if the utilization will be accordingly adjusted.

\subsection{Workloads}\label{ch5:ssec:workloads}
\begin{figure}[!t]
    \centering
    \begin{subfigure}{0.7\linewidth}
            \includegraphics[width=\textwidth]{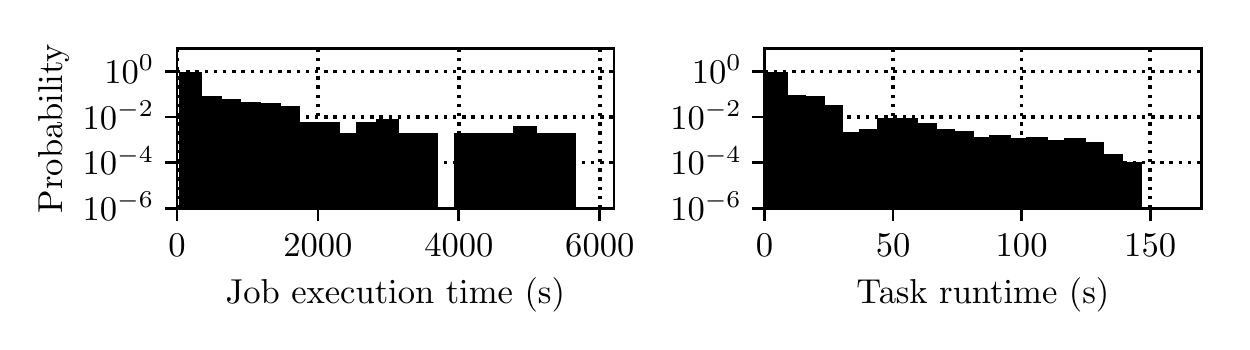}
            \caption{Workload~I.}
            \label{ch5:fig:workload2}
    \end{subfigure}
    \begin{subfigure}{0.7\linewidth}
            \includegraphics[width=\textwidth]{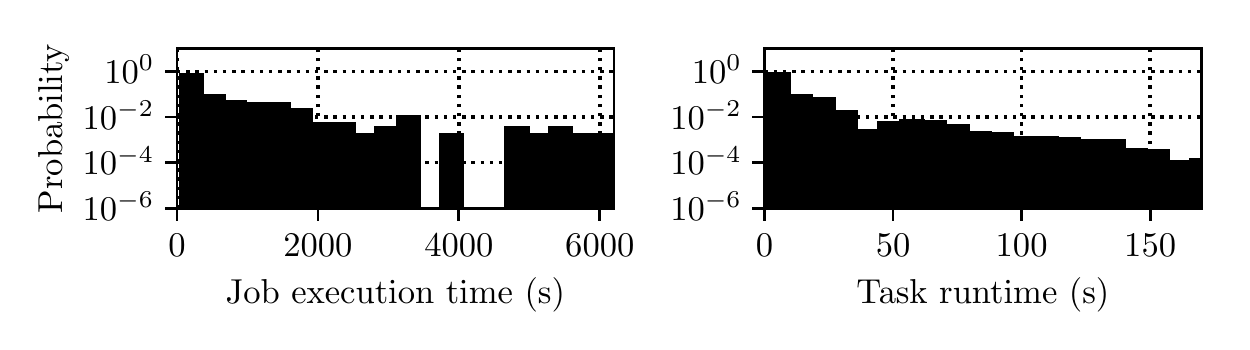}
            \caption{Workload~II.}
            \label{ch5:fig:workload2}
    \end{subfigure}
    \caption{Statistical characteristics of the workloads. The vertical axes have a log scale.}
    \label{ch5:fig:workloads}
\end{figure}

We use two workloads Workload~I and Workload~II,
each consisting of 600 workflows divided in three sets with 200 workflows each.
That allows us to perform three repetitions of each experiment.
Both workloads use the same 600 workflow structures, but differ in the task runtime characteristics.
We choose three popular scientific workflows from different fields, namely Montage, LIGO, and SIPHT.
The main reason for our choice is the existence of validated models for these workflow types.
Montage~\cite{jacob10} is used to build a mosaic image of the sky on the basis of smaller
images obtained from different telescopes. LIGO~\cite{abbott08} is used by the Laser
Interferometer Gravitational-Wave Observatory (LIGO) to detect gravitational waves. 
SIPHT~\cite{livny12} is a bioinformatics workflow used to discover bacterial regulatory RNAs.
We take the workflow structures, the task runtime distributions and file sizes
from the Bharathi generator~\cite{pegasuswfgenerator, bharathi08}.
We scale down the original task runtimes and file sizes
to reduce the total execution time of the workloads
by dividing the original values from the generator
by 30 and rounding them to the nearest integer.
We guarantee that the minimal task runtime is 1 second
and the minimal files size is 1KB.
Since we have two resource types in our model,
for each task we take its scaled down runtime
and generate one extra task runtime
using the uniform distribution.
For Workload~I the maximal deviation for the second task runtime
from the original task runtime is 50\%, and
the original and new task runtimes
are randomly assigned to the resource types.
For Workload~II the maximal deviation from
the original task runtime is 100\%,
and the new generated task runtime
is always assigned to the second resource type.
In both workloads each workflow has a randomly assigned priority in the range from 0 to 9.
Figure~\ref{ch5:fig:workloads} presents task runtime and job runtime distributions of the workload.
The details of each workload are summarized in Table~\ref{ch5:tab:workloads}.

\begin{table}[!t]
\centering
\caption{Characteristics of Workloads~I and~II.}
\label{ch5:tab:workloads}
\begin{threeparttable}
\begin{tabular}{lccc}
Property & WL~I & WL~II \\
\toprule
Total workflows in all three sets	&   \multicolumn{2}{c}{600} \\
Total tasks in all three sets	&	\multicolumn{2}{c}{44,340}	\\
Mean number of tasks in a workflow	&	\multicolumn{2}{c}{74}	\\
Median number of tasks in a workflow	&	\multicolumn{2}{c}{38}	\\
Standard deviation of number of tasks in a workflow & \multicolumn{2}{c}{95} \\
\hline					
Mean job execution time [s]	&	467	&	508	\\
Median job execution time [s]	&	276	&	303	\\
Standard deviation of job execution times [s]	&	692	&	761	\\
\hline					
Mean task runtime (averaged for both resource types) [s]	&	6.3	&	6.9	\\
Median task runtime (averaged for both resource types) [s]	&	1.5	&	1.5	\\
Standard deviation of task runtimes (averaged for both resource types) [s]	&	13.7	&	15.4	\\
Mean task runtime on the Small resource [s]	&	6.3	&	8.2	\\
Mean task runtime on the Large resource [s]	&	6.3	&	5.5	\\
Total task runtime (averaged for both resource types) [ks]	&	280	&	305	\\
\hline					
Mean task input data size [MB]	&	\multicolumn{2}{c}{578}	\\
Median task input data size [MB]    &	\multicolumn{2}{c}{138}	\\
Standard deviation task input data size [MB]	&	\multicolumn{2}{c}{1,364}	\\
Mean task output data size [MB]	&	\multicolumn{2}{c}{213}	\\
Median task output data size [MB]	&	\multicolumn{2}{c}{9}   \\
Standard deviation task output data size [MB]	&	\multicolumn{2}{c}{2,224}	\\
Total task input data size (including read duplicates\tnote{*}~) [TB]	&	\multicolumn{2}{c}{25,6}	\\
Total task output data size [TB]	&	\multicolumn{2}{c}{9,4}	\\
\bottomrule
\end{tabular}
  \begin{tablenotes}
    \item[*]When different tasks read the same file.
  \end{tablenotes}
\end{threeparttable}
\end{table}
\clearpage

%% file: sections/experimental-results.tex
\section{Experiment Results}\label{ch5:sec:experimental-results}
In this section, we present our experimental results.
We first analyze the runtimes of the considered autoscalers
obtained during the experiments.
Then we investigate how varying the budget affects
the workload performance and how it differs between the users.
Finally, we analyze the system-oriented, and elasticity metrics.
We report two experiment configurations, where we assign either equal budgets (eq.)
to both users or different budgets (diff.) for each user.
The sets of experiment configurations with regard
to the experiment results sections are summarized in Table~\ref{ch5:tab:experiment-configurations}.
Our results show that our PFA autoscaler
shows up to 76\% lower average algorithm runtime when
given the same workload as PLF and SCF,
while reducing by up to 47\% the average job slowdowns
The full set of software and computational artifacts used
to obtain the presented results are publicly available~\cite{ilyushkin2019pfa_software}, \cite{ilyushkin2019pfa_computational}.

\begin{table}[!t]\centering
\caption{Experiment configurations.}
\label{ch5:tab:experiment-configurations}
\begin{tabular}{lrrc}
Sec. & Budget Configuration & PFA Configuration & WL\\
\toprule
$\S$\ref{ch5:sec:workflow-slowdowns} & eq. 60\textcurrency, 80\textcurrency, 100\textcurrency, 120\textcurrency; diff. 120\textcurrency/80\textcurrency & $m$ = 10, 20, 30; $\alpha$ = 0.7, 0.8, 0.9 & I\\
$\S$\ref{ch5:sec:elasticity-performance} & eq. 60\textcurrency, 100\textcurrency, 120\textcurrency; diff. 120\textcurrency/80\textcurrency & $m$ = 10; $\alpha$ = 0.7 & I\\
$\S$\ref{ch5:sec:system-performance} & eq. 60\textcurrency, 100\textcurrency, 120\textcurrency; diff. 120\textcurrency/80\textcurrency & $m$ = 10; $\alpha$ = 0.7 & I\\
$\S$\ref{ch5:sec:autoscaling-dynamics} & eq. 120\textcurrency & $m$ = 10 & I \\
\hline
$\S$\ref{ch5:sec:workflow-slowdowns} & eq. 100\textcurrency & $m$ = 10; $\alpha$ = 0.7 & II\\

\bottomrule
\end{tabular}
\end{table}

\begin{figure}[!t]
\vspace{1.5em}
    \centering
    \includegraphics[width=0.6\linewidth]{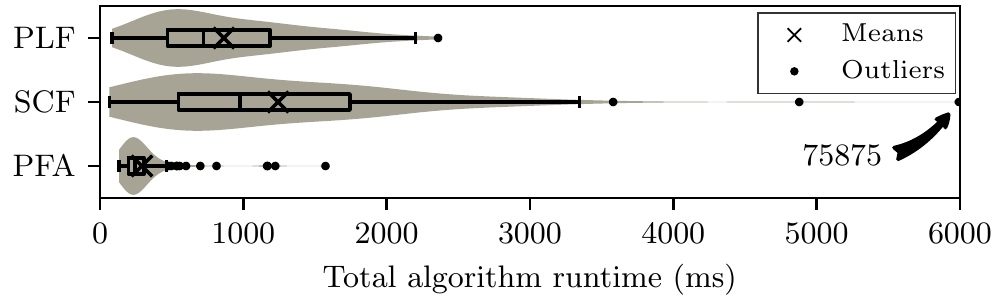}
    \caption{Variability of total runtimes for all the considered autoscalers.}
    \label{ch5:fig:algo-total-runtime}
\vspace{2em}
    \includegraphics[width=0.61\linewidth, trim={-1mm 0 8mm 0}]{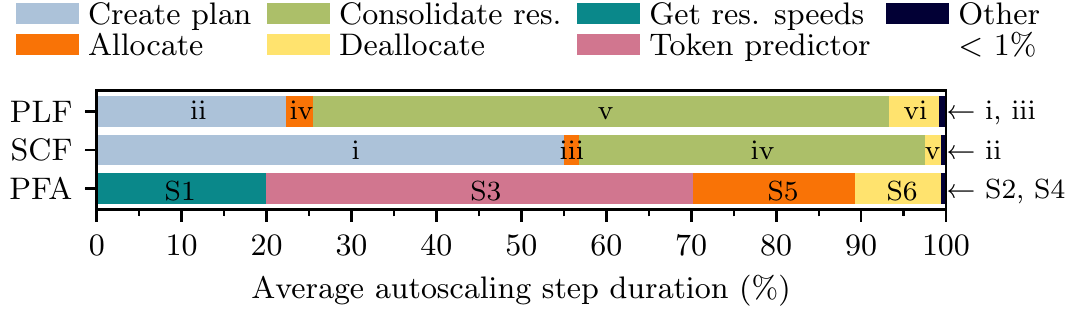}
    \caption{Average duration of each autoscaling step within the total algorithm runtime for each considered autoscaler.}
    \label{ch5:fig:algo-runtime-decomposition}
\end{figure}

\subsection{Algorithm Performance}\label{ch5:sec:algorithm-runtime}
Figure~\ref{ch5:fig:algo-total-runtime} shows the variability
of total algorithm runtimes executed at every autoscaler invocation.
The runtime of the algorithm varies depending on the number
of workflows that are currently in the system and depending on their characteristics.
We can also see that both plan-based autoscalers PLF and SCF have 3--4 times longer average
runtimes and show higher runtime variability than our PFA autoscaler.
Moreover, SCF autoscaler has one large outlier when it was running for 76 seconds, thus, exceeding the length of the autoscaling interval and
delaying the workload! Such behavior is very unfavourable
as it can negatively affect the stability of a WMS
during sudden demand surges.

Figure~\ref{ch5:fig:algo-runtime-decomposition} shows the average duration
of each autoscaling stage as a percentage of the average total algorithm runtime.
For PLF and SCF autoscalers the planning and
the resource consolidation steps take up 95\% of their
total execution time. Resource consolidation is basically
responsible for making the task placement plan.
For our PFA autoscaler the token-based
demand prediction takes on average
50\% of the total execution time.

\begin{figure}[!t]
    \centering
    \includegraphics[width=0.5\linewidth]{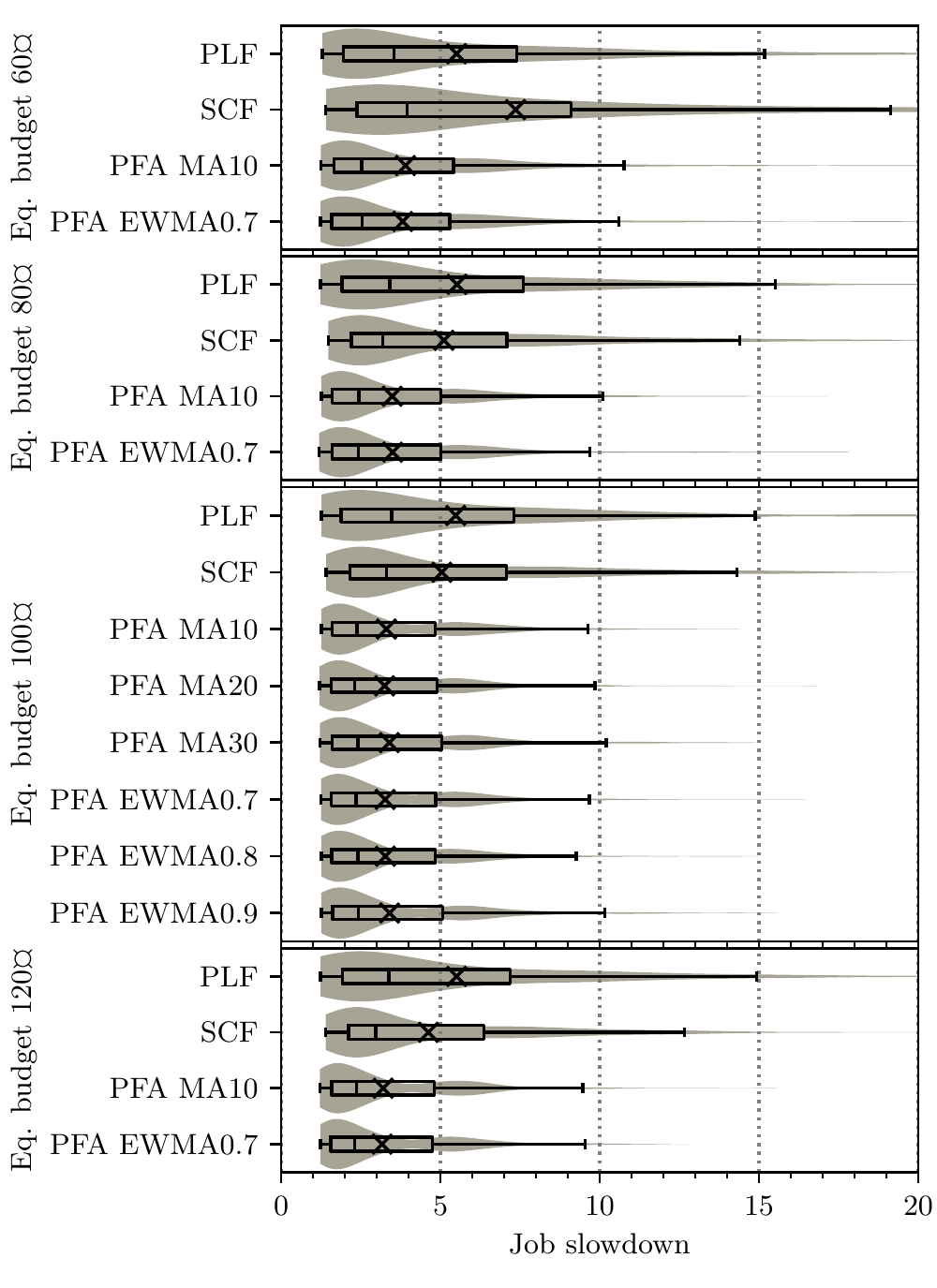}
    \caption{Variability of job slowdowns for all the studied autoscalers with equal budgets of 60\textcurrency, 80\textcurrency, 100\textcurrency, and 120\textcurrency~ when running WL~I. PFA autoscaler was executed with different smoothing methods. Means are marked with $\times$.}
    \label{ch5:fig:slowdowns-equal-budget}
\end{figure}

\subsection{Workload Performance}\label{ch5:sec:workflow-slowdowns}
Further, we analyze the job slowdowns to investigate
how the considered autoscalers affect the workload
performance from the end-user perspective.
Figure~\ref{ch5:fig:slowdowns-equal-budget} shows the variability of
job slowdowns in two configurations where both users are assigned with
equal budgets of 60\textcurrency, 80\textcurrency, 100\textcurrency, or 120\textcurrency, accordingly.
In this figure we can see a clear trend where higher budgets decrease the average job
slowdown as well as decrease the slowdown variability.
We use the configuration with equal budget 100\textcurrency~as a baseline,
as then each user can allocate at max. 52\% of the system resources.
With 100\textcurrency, the PFA policy is executed with various MA history depths $m$ of 10, 20, and 30,
and with EWMA $\alpha$ values of 0.7, 0.8, and 0.9.
We can observe that PFA in any of the considered configurations
shows lower average job slowdowns, as well as lower slowdown variability than PLF and SCF.
Different PFA smoothing methods do not significantly affect the PFA performance.

\begin{figure}[!t]
    \centering
    \includegraphics[width=0.5\linewidth]{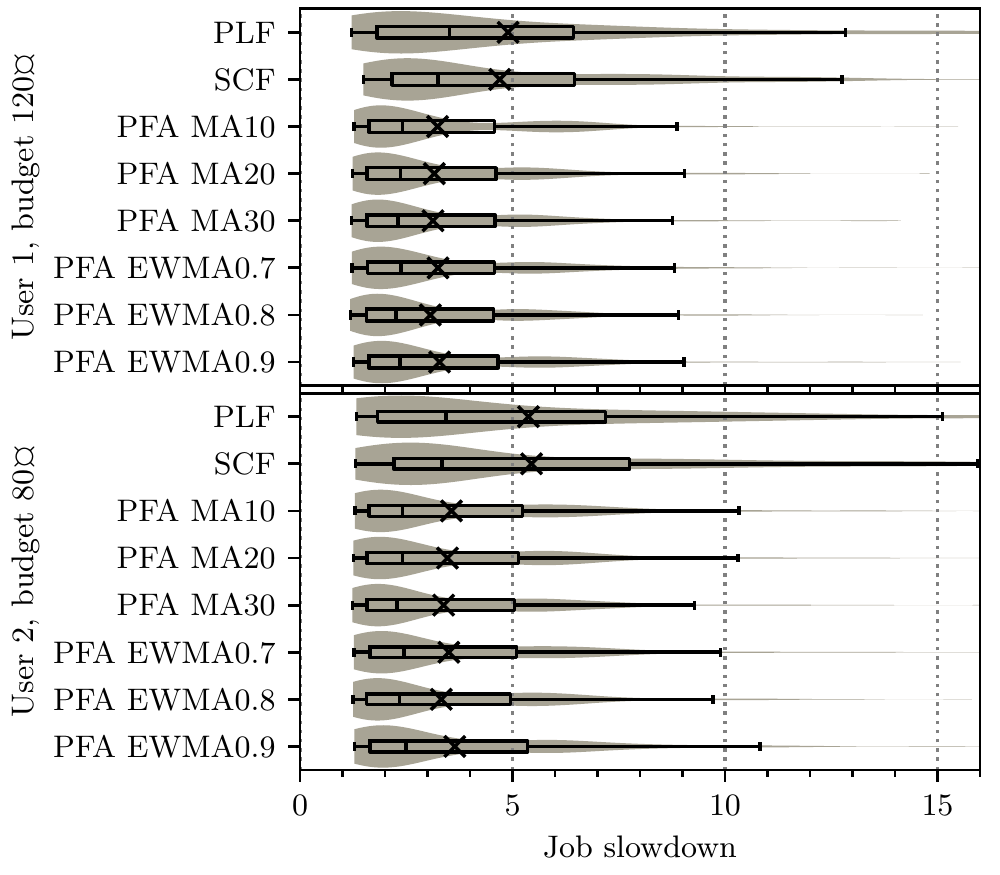}
    \caption{Variability of job slowdowns for all the studied autoscalers with different budgets for each user when running WL~I. PFA autoscaler was executed with different smoothing methods. Means are marked with $\times$.}
    \label{ch5:fig:slowdowns-different-budget}
\vspace{1.5em}
    \centering
    \includegraphics[width=0.5\linewidth]{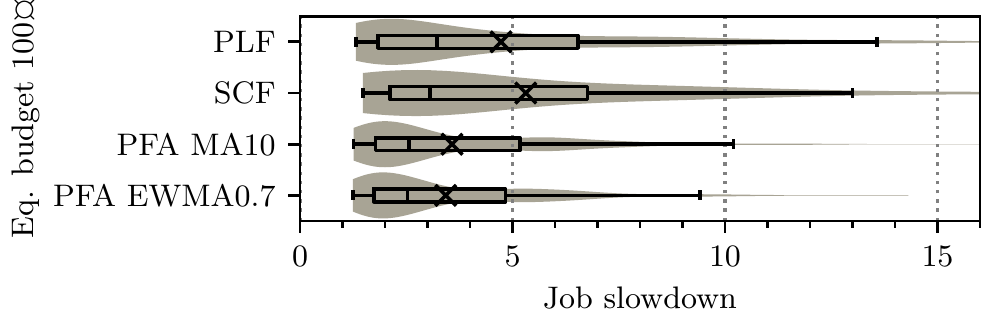}
    \caption{Variability of job slowdowns for all the studied autoscalers with equal budgets for both users when running WL~II. PFA autoscaler was executed with MA depth 10 and EWMA with pole value 0.7. Means are marked with $\times$.}
    \label{ch5:fig:slowdowns-equal-budget-wl2}
\end{figure}

Figure~\ref{ch5:fig:slowdowns-different-budget} shows job slowdown variability
for the configuration where User~1 has higher budget 120\textcurrency~than
User~2 with budget 80\textcurrency. We can conclude, that all the considered
autoscalers guarantee that the user with the higher budget gets better performance, since User~1 has lower average job slowdowns and lower slowdown variability.
PFA autoscaler for both users shows better workload performance than PLF and SCF.

Figure~\ref{ch5:fig:slowdowns-equal-budget-wl2} presents job slowdowns
for configuration with equal budget 100\textcurrency~for Workload~II.
The observed trend is the same as in Figures~\ref{ch5:fig:slowdowns-equal-budget} and \ref{ch5:fig:slowdowns-different-budget}.
The tasks in WL~II
on average run faster on the Large resource type than on the Small resource type (see Table~\ref{ch5:tab:workloads}). Thus, we can conclude,
that all the considered autoscalers can successfully operate
with workloads where tasks ``prefer'' a specific resource type.

\begin{figure}[!t]
   \begin{minipage}{0.48\textwidth}
            \centering
            \includegraphics[width=\linewidth]{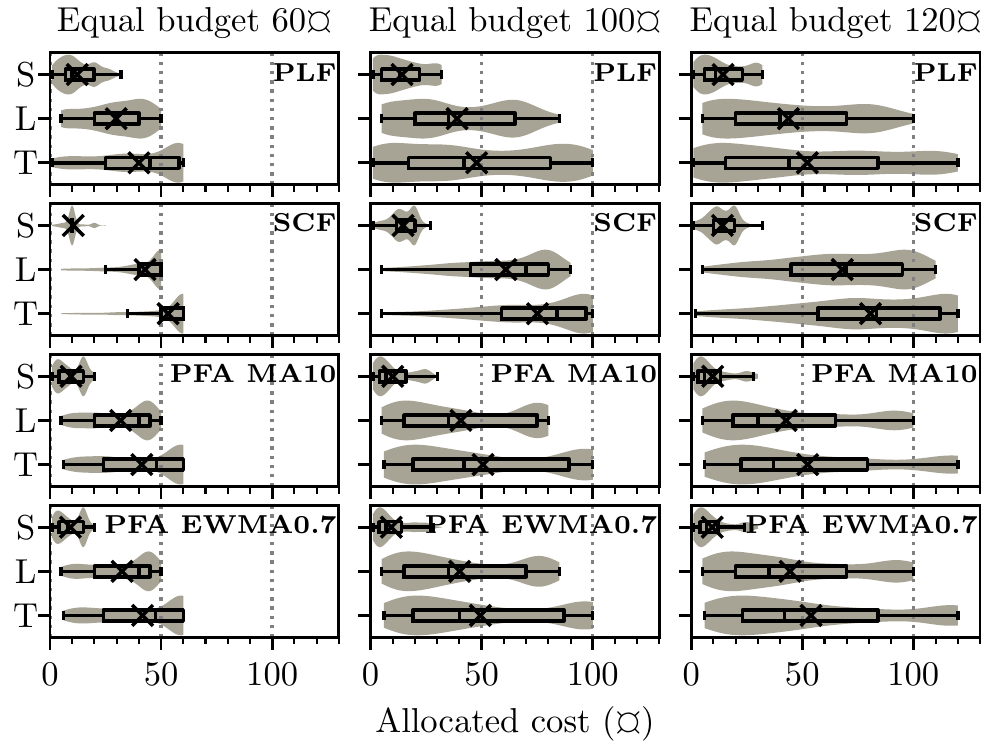}
            \caption{Variability of monetary cost for allocated resources
            per billing interval for User~1 for the studied autoscalers with
            equal budgets of 60\textcurrency, 100\textcurrency, and 120\textcurrency~when running WL~I. For Small and Large resource types, and in Total. Means are marked with $\times$.}
            \label{ch5:fig:monetary-cost-equal-budget}
   \end{minipage}\hfill
   \begin{minipage}{0.48\textwidth}
            \centering
            \includegraphics[width=\linewidth,trim={0 0 0 6mm}]{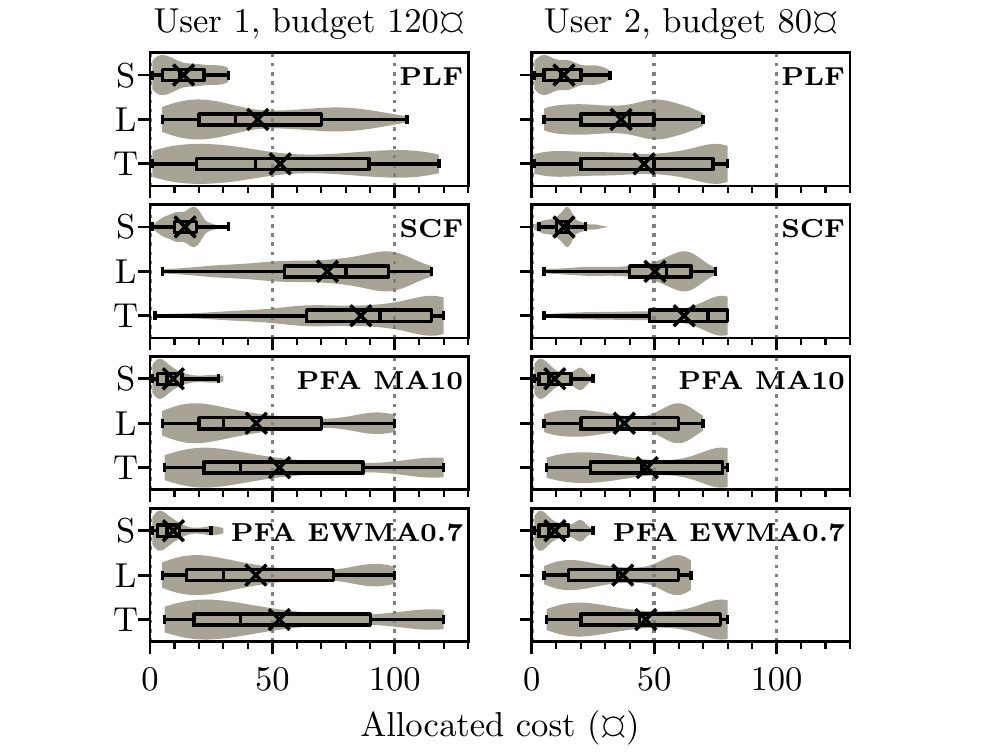}
            \caption{Variability of monetary cost for allocated resources
            per billing interval for each user for the studied autoscalers with
            different budgets when running WL~I.
            For Small and Large resource types, and in Total.
            Means are marked with $\times$.}
            \label{ch5:fig:monetary-cost-diff-budget}
   \end{minipage}

\vspace{1.5em}
   \begin{minipage}{0.48\textwidth}
            \centering
            \includegraphics[width=\linewidth]{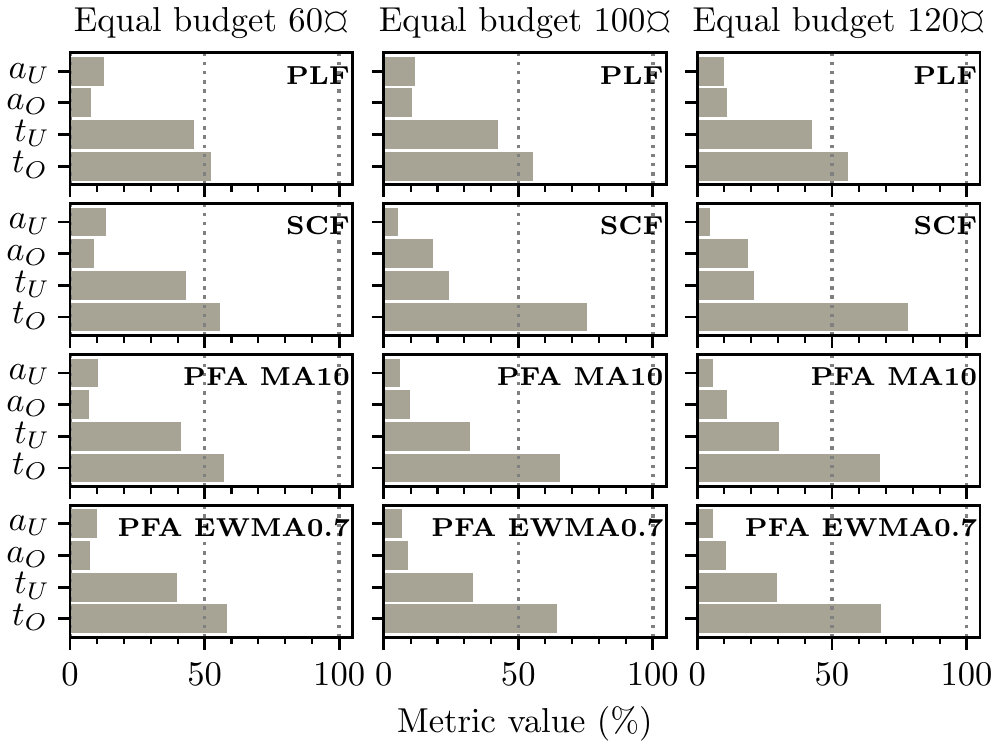}
            \caption{Elasticity metrics for User~1 for the studied autoscalers with equal budgets of 60\textcurrency, 100\textcurrency, 120\textcurrency~when running WL~I.
            For all the metrics lower values are better.}
            \label{ch5:fig:as-metrics-equal-budgets}
   \end{minipage}\hfill
   \begin{minipage}{0.48\textwidth}
            \centering
            \includegraphics[width=\linewidth]{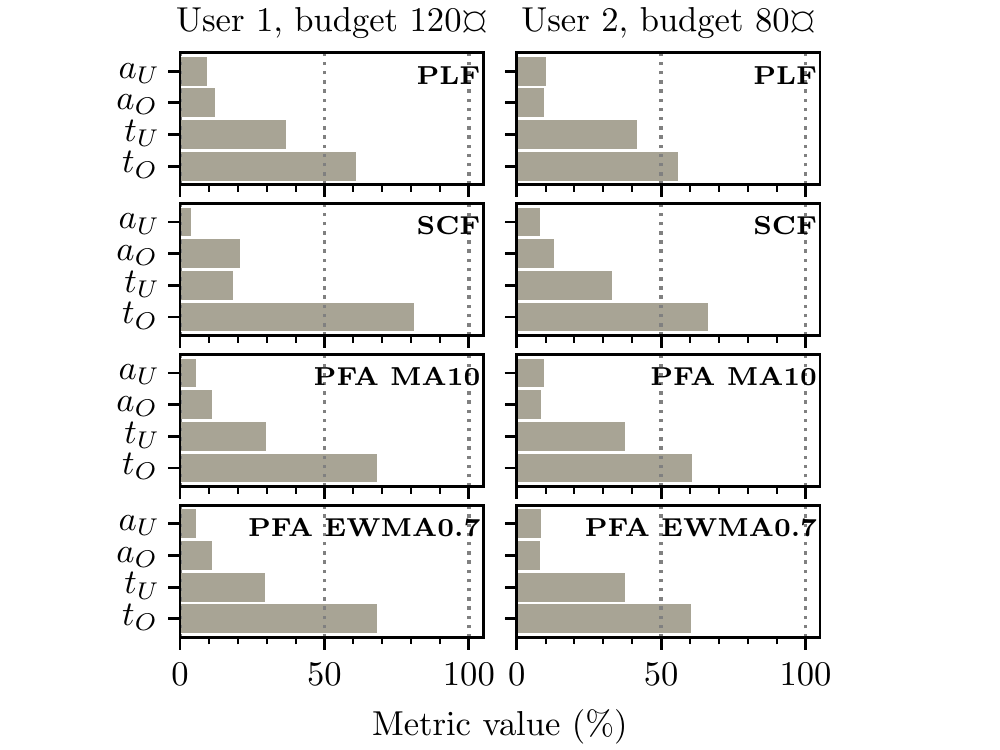}
            \caption{Elasticity metrics for each user for the studied autoscalers with different budgets 120\textcurrency~and 80\textcurrency~when running WL~I.
            For all the metrics lower values are better.}
            \label{ch5:fig:as-metrics-different-budgets}
   \end{minipage}
\end{figure}

From Figure~\ref{ch5:fig:monetary-cost-equal-budget} 
and Figure~\ref{ch5:fig:monetary-cost-diff-budget} 
we can
see that no autoscalers exceed the budget constraint
for the configurations with equal and different budgets.
SCF on average spends more budget than PLF and PFA.
Our PFA autoscaler shows comparable mean costs to PLF,
but lower median costs at higher budgets.
Moreover, for PFA, when the budget is large enough, the distribution
of allocated costs skews towards lower values.
For all the autoscalers most of the cost comes from
the Large resource type, as it is more expensive.
Further, when presenting the results, we do not plot
some experiment configurations
if these configurations
show no significant difference.
E.g., from Figure~\ref{ch5:fig:monetary-cost-equal-budget} we omit
the results with the equal budget 80\textcurrency, and from Figure~\ref{ch5:fig:monetary-cost-diff-budget} we omit
the results for PFA with the configurations MA $m$ = 20, 30,
and $\alpha$ = 0.8, 0.9.

\vspace{-1em}
\subsection{Elasticity Performance}\label{ch5:sec:elasticity-performance}
\vspace{-0.5em}
Figure~\ref{ch5:fig:as-metrics-equal-budgets} shows the considered elasticity metrics
for the configurations with equal budget of 60\textcurrency, 100\textcurrency, and 120\textcurrency~for User~1. We do not report values
for User~2, as we do not observe significant difference between the users.
Figure~\ref{ch5:fig:as-metrics-different-budgets} shows elasticity metrics
for both user for the configuration with different budgets of
120\textcurrency~and 80\textcurrency~for User~1 and User~2, accordingly.
When calculating the elasticity metrics, we skip the periods where
demand exceeds the maximal resource number of 64.
The resource demand can vary significantly even at relatively
low utilization of 20\% as it depends on the structure and LoP~of the workflows.

In Figure~\ref{ch5:fig:as-metrics-equal-budgets} we can see that
for budgets 100\textcurrency~and 120\textcurrency,
SCF shows in all the plots the worst values for $a_O$ and $t_O$,
it also has the best values for $a_U$ and $t_U$.
In other words, SCF tends to over-provision.
For example, in the configuration with budget 100\textcurrency, SCF over-provisions
for almost 75\% of the time with on average 18\% too many resources
and has in 24\% of the time on average 5\% too few resources.
At lower budget 60\textcurrency, SCF spends less time over-provisioning,
but still shows on average the worst over-provisioning accuracy of 8.5\%.

In contrast, PLF with budget 100\textcurrency~has in 42\% of the time
on average 11\% too few resources. Thus, PLF tends to under-provision the system and has the worst values for $a_u$ (except for budget 60\textcurrency, where SCF is the worst), $t_U$ and only the best values for $t_O$.

Our PFA autoscaler shows the best values for $a_O$.
Further, PFA has the second best values for $a_U$ with budgets 100\textcurrency~and 120\textcurrency. For budget 60\textcurrency~PFA
shows the best value for $a_U$, $t_U$, but the words value for $t_O$, which is, however, compensated by low $a_O$.
In general, PFA is more accurate than the other two autoscalers,
as it has the lowest summed up $a_U$ and $a_O$ accuracy values.
Moreover, spending more time under- or over-provisioning with higher accuracy is more favourable
than spending less time under- or over-provisioning with lower accuracy.
The same trends can be observed for the configuration with different budgets in Figure~\ref{ch5:fig:as-metrics-different-budgets}.
From this we conclude that our approach is more likely to satisfy the user SLOs,
which is also confirmed by the workload performance results.
Although, PFA does not use known in advance task runtime estimates,
it is more accurate when applied to workload of workflows than the plan-based autoscalers.

\begin{figure}[!tb]
   \begin{minipage}{0.48\textwidth}
            \centering
            \includegraphics[width=\linewidth]{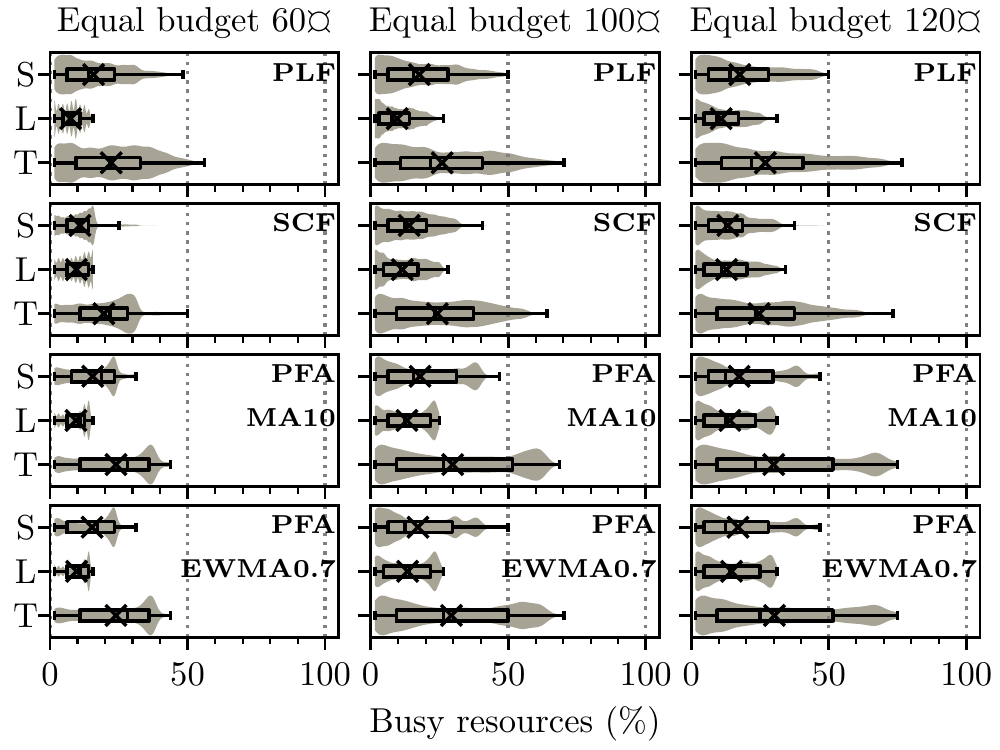}
            \caption{Variability of busy resources for User~1 for the studied autoscalers with equal budgets of 60\textcurrency, 100\textcurrency, 120\textcurrency~when running WL~I.
            For Small and Large resource types, and in Total.
            Means are marked with $\times$.}
            \label{ch5:fig:busy-resources-equal-budget}
   \end{minipage}\hfill
   \begin{minipage}{0.48\textwidth}
            \centering
            \includegraphics[width=\linewidth,trim={0 0 0 6mm}]{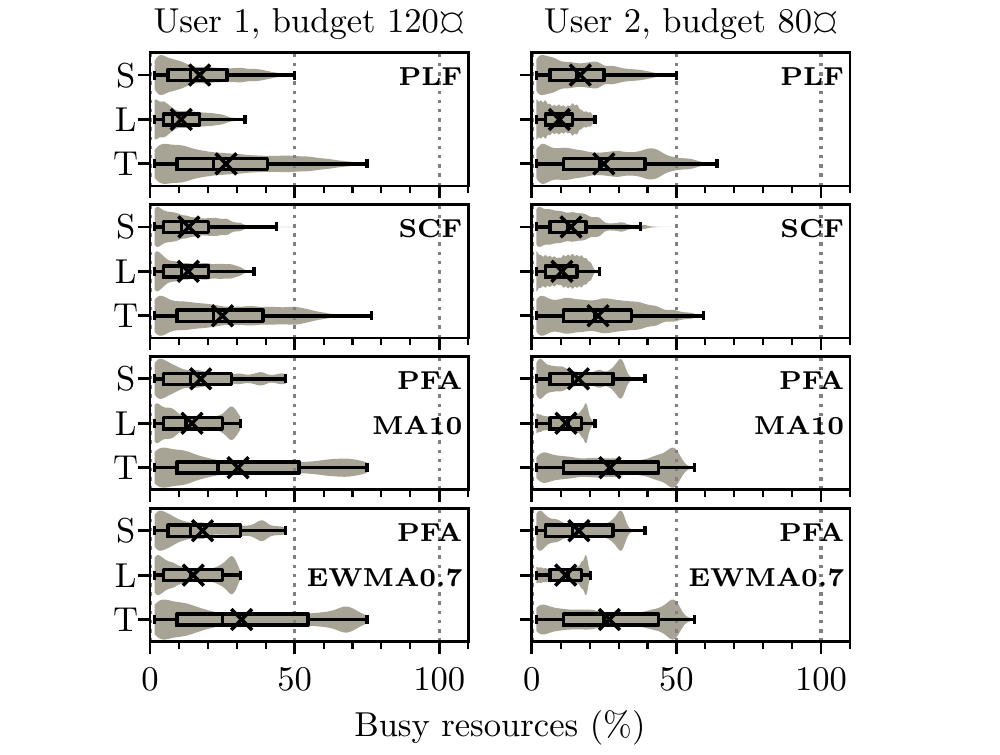}
            \caption{Variability of busy resources for each user
             with different budgets 120\textcurrency~and 80\textcurrency~when running WL~I. For Small and Large resource types, and in Total.
            Means are marked with $\times$.}
            \label{ch5:fig:busy-resources-different-budget}
   \end{minipage}

\vspace{1.5em}
   \begin{minipage}{0.48\textwidth}
            \centering
            \includegraphics[width=\linewidth]{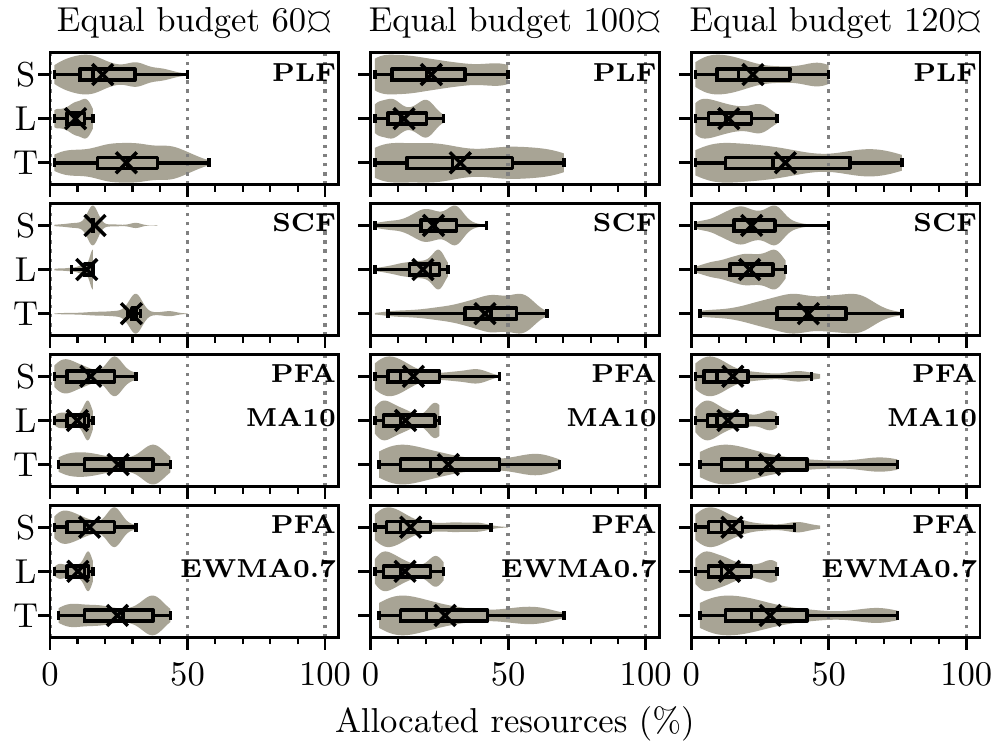}
            \caption{Variability of allocated resources for User~1 for the studied autoscalers with equal budgets of 60\textcurrency, 100\textcurrency, 120\textcurrency~when running WL~I.
            For Small and Large resource types, and in Total.
            Means are marked with $\times$.}
            \label{ch5:fig:allocated-resources-equal-budget}
   \end{minipage}\hfill
   \begin{minipage}{0.48\textwidth}
            \centering
            \includegraphics[width=\linewidth,trim={0 0 0 0}]{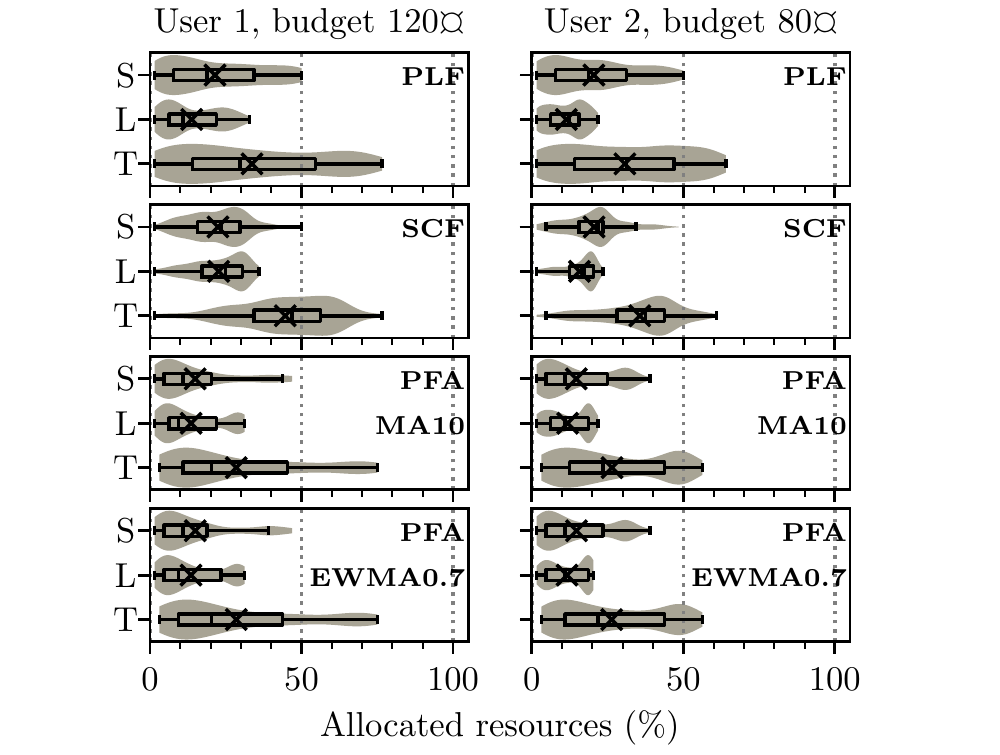}
            \caption{Variability of allocated resources for each user
             with different budgets 120\textcurrency~and 80\textcurrency~when running WL~I. For Small and Large resource types, and in Total.
            Means are marked with~$\times$.}
            \label{ch5:fig:allocated-resources-different-budget}
   \end{minipage}
\end{figure}

\subsection{System-Oriented Performance}\label{ch5:sec:system-performance}
We look at the percentage of busy and allocated
resources throughout the experiment
to evaluate the system-oriented performance, as these metrics
shows how effectively the resources are utilized,
and how many resources are actually allocated.
Figure~\ref{ch5:fig:busy-resources-equal-budget} presents the percentage of busy resources 
for the configurations with equal budget of 60\textcurrency, 100\textcurrency, and 120\textcurrency~for User~1. Figure~\ref{ch5:fig:allocated-resources-equal-budget} shows the variability of allocated resources for the same configuration and the same user.
We do not report values
for User~2, as we do not observe significant difference between the users.

Figure~\ref{ch5:fig:busy-resources-different-budget} shows the percentage of busy resources for both user for the configuration with different budgets of
120\textcurrency~and 80\textcurrency~for User~1 and User~2, accordingly.
Figure~\ref{ch5:fig:allocated-resources-different-budget} shows the percentage of allocated resources for different budgets also for both users.

SCF shows lowest average number of busy resources, which correlates with
the elasticity results, as SCF tends to over-provision more.
PFA shows higher and also more balanced use of the resources.
We can also see that the variability of busy resources
increases together with the budget.
Looking at the variability of allocated resources, we observe that PLF and SCF
on average allocate more resources than PFA, this correlates with the results on monetary costs from Section~\ref{ch5:sec:workflow-slowdowns}.
For higher budgets PFA tends to spend more time allocating less resources than the other two autoscalers. For lower budgets the difference between the autoscalers decreases. Thus, we can conclude, that,
in contrast to PLF and SCF, PFA
allocates and uses the resources more efficiently,
while given the same budget.

\begin{figure}[!t]
    \centering
    \includegraphics[width=0.9\linewidth]{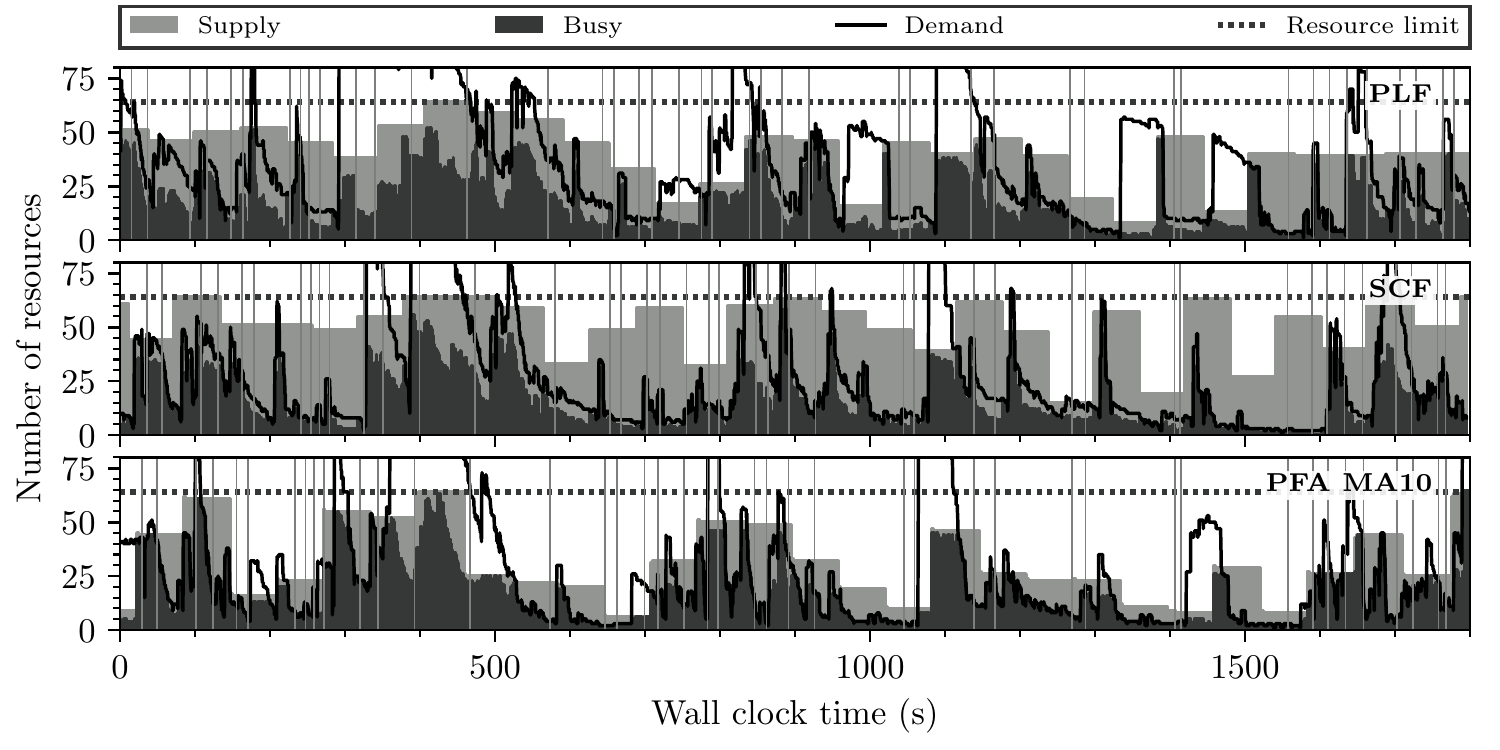}
    \caption{The dynamics of autoscaling on a cropped interval of 1,800 seconds for both users
    with equal budget 120\textcurrency. Vertical lines indicate workflow arrivals.}
    \label{ch5:fig:dynamics}
\end{figure}

\vspace{-1em}
\subsection{Autoscaling Dynamics}\label{ch5:sec:autoscaling-dynamics}
\vspace{-0.5em}
We further study the dynamics of the obtained Airflow traces
to better understand the performance differences between the autoscalers.
Figure~\ref{ch5:fig:dynamics} shows
the snapshots of autoscaling dynamics on a cropped
interval of 1,800 seconds for both users with equal budget of 120\textcurrency.
We rely on the configuration with 120\textcurrency~as it shows higher supply variability.
We can see that PLF and SCF autoscalers
have higher demand values---the number of waiting eligible tasks.
Both PLF and SCF show lower resource utilization (the number of busy resources)
as in between the autoscaler invocations the tasks are waiting
for being included in the plan.
Moreover, we can see how PLF makes wrong predictions, e.g.,
at the time around 1350 seconds, as PLF makes its predictions
using the tasks that are eligible at the moment it is invoked.
Similar-looking spikes can be observed for PFA, e.g., at the time
around 1425 seconds, however, it is caused by a double workflow
arrival within the autoscaling interval.
We can also observe different shapes of demand curves in each plot, as
the number of eligible tasks depends on the throughput, that, in turn,
depends on the number of allocated resources and the efficiency
of the task placement policy utilized by the scheduler.
Interestingly, for PFA on the interval 350--450 seconds the
allocated resources are not fully utilized, even though the demand
exceeds the resource ceiling. The reason for this is the latency caused
by the Airflow system.

The phases of the autoscaling intervals are slightly drifting
between the plots, as we did not set a goal to deliberately synchronize the phases of different autoscalers. The drifting is caused by possible internal delays in the Airflow system during the demand surges, and by occasional delays in the autoscalers, e.g., when SCF runs for too long, as shown in Figure~\ref{ch5:fig:algo-total-runtime}.
That is why, to minimize the possible effect of such drifting,
we run three independent subsets of workflows within the workload.

%% file: sections/optimal-solution.tex
\section{The Optimal Solution}\label{ch5:sec:optimal-solution}
In this section, to validate the performance of the considered policies,
we compare the results obtained from the Airflow system with
the optimal solution obtained from solving the optimization problem represented
as a Mixed Integer Programming (MIP) model.
For that we modify the MIP model proposed
by Wang et al.~\cite{wang2017using}
to incorporate budget constraints,
while following the similar notation,
and implement the model in the Gurobi~\cite{gurobi} solver (v.~8.0.1).
The goal of the solver is to find the optimal plan which, under
the budget and resource constraints, finds the task placement plan and
determines the number of resources of each type
that should be allocated on each autoscaling interval,
so that the response time of each workflow is minimized.

\subsection{Mixed Integer Programming Model}\label{ch5:ssec:mip-model}
\begin{table}[!t]
\centering
\caption{Symbols used for the MIP model.}\label{ch5:tab:mip-symbols}
\begin{tabularx}{\linewidth}{lX}
\toprule
$\mathcal{T}$ & The set of time slots $\mathcal{T} = \{1, 2, \ldots, T\}$\\
$\mathcal{W}$ & The set of workflows $\mathcal{W} = \{1, 2, \ldots, W\}$\\
$\mathcal{S}$ & The set of workflows tasks $\mathcal{S} = \{1, 2, \ldots, S\}$\\
$\mathcal{M}$ & The set of billing intervals $\mathcal{M} = \{1, 2, \ldots, M\}$\\
$\mathcal{V}$ & The set of resources $\mathcal{V} = \{1, 2, \ldots, V\}$\\
$L$ & Number of time slots per billing interval\\
$B$ & Budget per billing interval\\
$A_w$ & Arrival time of workflow $w$\\
$D_w$ & Earliest possible completion time for workflow $w$\\
$R_{j,k}$ & Runtime of task $j$ on resource $k$\\
$P_k$ & Cost of running resource $k$ on a billing interval\\
$l_{i,j}$ & Equals one if task $i$ depends on task $j$\\
$h_w(t)$ & The value of workflow $w$ finishing at time $t$\\
\hline
$x_{j,k}^t$ & Binary variable, equals one if task $j$ starts at time $t$ on resource $k$\\
$y_k^m$ & Binary variable, equals one if resource $k$ is active on billing interval $m$\\
$z_k^m$ & Integer variable, determines the number of active time slots for resource $k$ on billing interval $m$\\
$u_w^t$ & Binary variable, equals one if workflow $w$ finishes at time $t$\\
\bottomrule
\end{tabularx}
\end{table}

The MIP model presents time as a set $\mathcal{T} = \{1, 2, \ldots, T\}$
of discrete time slots of equal duration, where $T$ is the furthest time horizon.
The times slots are grouped into $M$ billing intervals.
Each billing interval consists of $L$ time slots.
The set of billing intervals is denoted by $\mathcal{M} = \{1, 2, \ldots, M\}$,
and $T$ is divisible by $L$, so that $M =T / L$.
The budget $B$ is given per billing interval and should not be exceeded.
The input of the problem is a set of workflows $\mathcal{W} = \{1, 2, \ldots, W\}$,
where each workflows contain tasks. All the tasks in all the workflows are represented
by the set $\mathcal{S} = \{1, 2, \ldots, S\}$, where each task can belong to a single workflow only.
The task precedence constraints are represented by a binary matrix $(l_{i,j}), \forall i,j \in S$,
where $l_{i,j} = 1$ if task $i$ depends on task $j$, i.e., $i$ can start only
after $j$ has finished, and $l_{i,j} = 0$ otherwise. By convention, each $l_{i,i} = 0$.
Each workflow $w$ has an arrival time $A_w$, known in advance length
$B_w$ of its critical path, and earliest possible completion time $D_w$, so that $D_w = A_w + B_w - 1$.
The model also defines a set of computing resources $\mathcal{V} = \{1, 2, \ldots, V\}$.

If a task is scheduled on a resource, it runs on it exclusively until completion.
To represent the task assignment, we use binary decision variables $x_{j,k}^t$,
where $x_{j,k}^t = 1$ if task $i$ is scheduled to run on resource $k$ starting at time slot $t$,
and $x_{j,k}^t = 0$ otherwise. Each task should start only once, which we specify as follows:
\begin{equation} \label{ch5:eq:constr1}
    \sum_{k \in \mathcal{V}} \sum_{t \in \mathcal{T}} x_{j,k}^t = 1, \forall j \in \mathcal{S}.
\end{equation}

Let the integer variable $0 \leq z_k^m \leq L$ denote the number of active
time slots on each resource $k$ on billing interval $m$.
This requires the following constraints:
\begin{equation} \label{ch5:eq:constr2}
\begin{split}
    \sum_{t = (m - 1) \cdot L + 1}^{m \cdot L} \sum_{j \in S} \sum_{r=\max(1, t-R_{j,k}+1)}^{t} x_{j,k}^r = z_k^m, 
    \forall k \in \mathcal{V}, \forall m \in \mathcal{M}.
\end{split}
\end{equation}
Let the binary variable $y_k^m$ denote the active/idle state of each resource $k$ on billing interval $m$,
with $y_k^m = 1$ if some tasks are assigned on the resource, and $y_k^m = 0$ otherwise.
If the resource has no tasks scheduled, it is considered deallocated,
however if even a single task is assigned to the resource, it is considered active.
Accordingly, we define the following constraints:
\begin{equation} \label{ch5:eq:constr3}
\begin{split}
    y_k^m = \min(1, z_k^m), \forall k \in \mathcal{V}, \forall m \in \mathcal{M}.
\end{split}
\end{equation}

The tasks are not allowed to overlap, i.e., for each time slot and each resource at most
one task is allowed to occupy the time slot on that resource.
Let $R_{j,k}$ denote the running time of task $j$ on resource $k$ which is known in advance.
The non-overlapping constraints are specified as follows:
\begin{equation} \label{ch5:eq:constr4}
    \sum_{j \in \mathcal{S}} \sum_{r = \max(1, t-R_{i,k} + 1)}^{t} x_{j,k}^r \leq 1, \forall k \in \mathcal{V}, \forall t \in \mathcal{T}.
\end{equation}

The precedence constraints are formulated as follows:
\begin{equation} \label{ch5:eq:constr5}
\begin{split}
    \Big( \sum_{k \in \mathcal{V}} \sum_{t \in \mathcal{T}} t \cdot x_{i,k}^t - \sum_{k \in \mathcal{V}} \sum_{t \in \mathcal{T}} (t + R_{j,k}) \cdot x_{j,k}^t \Big) \cdot l_{i,j} \geq 0, \\
    \forall i,j \in \mathcal{S}.
\end{split}
\end{equation}

Further we formulate the constraints that no tasks of any workflow
can be scheduled to start before its arrival time:
\begin{equation} \label{ch5:eq:constr6}
    \sum_{k \in \mathcal{V}} \sum_{t \in \mathcal{T}} t \cdot x_{j,k}^t \geq A_w, \forall w \in \mathcal{W}, \forall j \in w.
\end{equation}

Since the optimization goal is to minimize the workflow response time within the given budget,
we represent it as a profit maximization problem where higher profit corresponds to a shorter response time.
For that, let $h_w : \{1,2,\ldots\} \to \mathbb{R}$ be a non-increasing value function,
where $h_w(t)$ represents the value gained depending on the time slot $t$ where the workflow $w$ is finished:
\begin{equation}
    h_w(t) = 
    \begin{cases}
    1, & \text{if } t \leq D_w,\\
    D_w - t, & \text{otherwise}.
    \end{cases}
\end{equation}
For each workflow $w$ and each time $t$ we define a binary variable $u_w^t$,
where $u_w^t = 1$ if workflow $w$ is completed at time $t$.
Since each workflow can finish only once, we formulate the following constraints:
\begin{equation} \label{ch5:eq:constr7}
    \sum_{t \in \mathcal{T}} u_w^t = 1, \forall w \in \mathcal{W}.
\end{equation}
The completion time of the workflow can be written as 
$\sum_{t \in \mathcal{T}} t \cdot u_w^t$.
Accordingly, the constraint that all the tasks of a workflow $w$ are
completed by the workflow completion time can be formulated as follows:
\begin{equation} \label{ch5:eq:constr8}
\begin{split}
    \sum_{k \in \mathcal{V}} \sum_{t \in \mathcal{T}} (t + R_{j,k} - 1) \cdot x_{j,k}^t \leq \sum_{t \in \mathcal{T}} t \cdot u_w^t, \forall w \in \mathcal{W}, \forall j \in w.
\end{split}
\end{equation}
Let $P_k$ be the cost of resource $k$, then
the budget constraints are defined as follows:
\begin{equation} \label{ch5:eq:constr9}
  \sum_{k \in \mathcal{V}} P_k \cdot y_k^m \leq B, \forall m \in \mathcal{M}.
\end{equation}
Finally, we can formulate the profit maximization objective:
\begin{equation} \label{ch5:eq:objective}
\begin{split}
&    \max \sum_{w \in \mathcal{W}} \sum_{t \in \mathcal{T}} h_w(t) \cdot u_w^t\\
&    \text{s.t. (\ref{ch5:eq:constr1})(\ref{ch5:eq:constr2})(\ref{ch5:eq:constr3})(\ref{ch5:eq:constr4})(\ref{ch5:eq:constr5})(\ref{ch5:eq:constr6})(\ref{ch5:eq:constr7})(\ref{ch5:eq:constr8})
(\ref{ch5:eq:constr9})}. 
\end{split}
\end{equation}

\subsection{Heuristics vs. the Optimal Solution}\label{ch5:ssec:mip-results}
We use three subsets of five workflows each from Workload~I, 
submitted with a fixed interval of 30 seconds in a system with
16 resources (vs. 64 resources in other experiments)
of two types with 8 resources in each.
For this reason, the maximal budget required to allocate all the system resources is 48\textcurrency.
The first group of five workflows 
consists of one Montage, one SIPHT, and three LIGO workflows,
with 183 tasks in total. The second group contains one Montage, two SIPHT and two LIGO workflows,
with 241 task in total. The third group contains two Montage, one SIPHT,
and two LIGO workflows with 199 tasks in total.
Further, we refer to these 15 workflows as the MIP workload.
We limit the number of considered workflows and limit the number of resources
due to much higher expected computational effort for finding the optimal solution
versus the considered heuristic approaches.
To make the workflows compatible with the MIP model, we round their task
runtimes to 5 seconds, which is the duration of the time slot in the MIP model,
and set the sizes of all the exchanged files to zero to make the comparison more fair.
We configure the MIP model with 16, 18, or 19 billing intervals (depending on the workload subset) and set the length of each billing interval to 3 time slots.
Accordingly, we set the Airflow autoscaling interval to 15 seconds.
We use a single user only, and find the optimal solutions with three different budgets of 10\textcurrency, 30\textcurrency, and 40\textcurrency.

\begin{figure}[t!]
    \centering
    \includegraphics[width=0.6\linewidth]{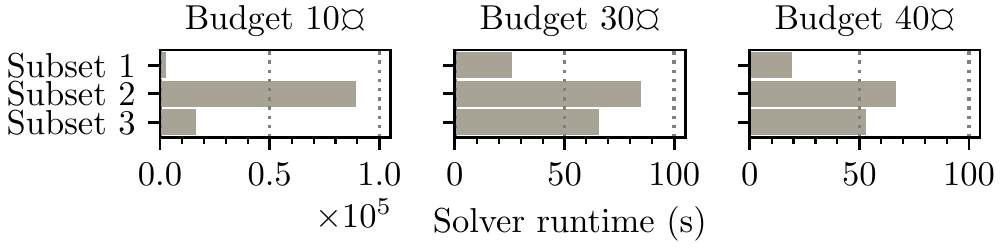}
    \caption{Solver runtimes for all three subsets of workflows with different budget constraints. In the left plot the x axis has much larger scale than in the other two plots.}
    \label{ch5:fig:mip-runtimes}
\end{figure}

\begin{figure}[t!]
    \centering
    \includegraphics[width=0.6\linewidth]{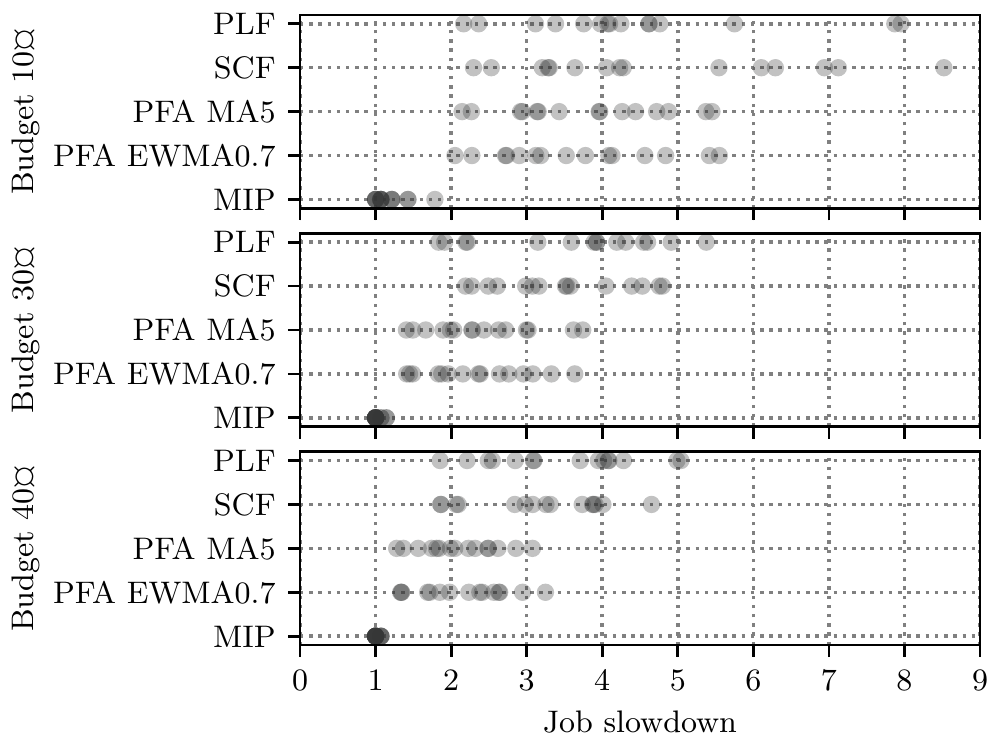}
    \caption{The variability of job slowdown for different budgets for Airflow-based and MIP results.}
    \label{ch5:fig:optimal-results-slowdowns}
\end{figure}

Figure~\ref{ch5:fig:mip-runtimes} shows solver runtimes for all three groups
of workflows with different budget constraints.
Note, that the lower the budget constraint the more time the solution takes.
For budget 10\textcurrency~finding the solution takes up to 88,592 seconds (24.5 hours)!
This confirms that the solution time does not scale linearly as it highly depends
on parameterization of the model, for example, on the chosen number of slots.
Moreover, the Gurobi solver has more than 40 MIP-related internal parameters~\cite{gurobiparameters} that
can significantly affect the performance of the solver.
To somehow automate this process, Gurobi even provides a parameter tuning tool~\cite{gurobituning}.
Since in our MIP model we add the budget constraints, they increase
the total runtime for finding the optimal solution, compared to
the runtimes reported
in the paper by Wang et al.~\cite{wang2017using}.
Even in the paper by Wang et al. the number of considered workflows
used with the MIP solver was much higher (500),
those workflow structures were very simple, and the model did not have budget constraints.
From this we can conclude that the MIP approach is not suitable for autoscaling workloads of workflows.

In Figure~\ref{ch5:fig:optimal-results-slowdowns} we
compare the slowdowns of the workflows from the optimal plan
with the slowdown obtained from running the same workflows
in our Airflow setup with all the three considered autoscalers.
We configure PFA with $m = 5$ for MA due to the low number
of autoscaling intervals in the MIP model,
for EWMA we use $\alpha = 0.7$.
We do not use violin plots in Figure~\ref{ch5:fig:optimal-results-slowdowns} for the distributions, as for each run there we have only 15 samples---the total number of executed workflows in all the three MIP workload subsets.
We can clearly see that the slowdowns obtained from the Airflow system are up to 8 times higher than the slowdowns from the MIP solution. Note, that the slowdowns from the Airflow experiments also include the slowdown caused by the WMS itself. However,
the general trend is similar to Figures~\ref{ch5:fig:slowdowns-equal-budget}, and \ref{ch5:fig:slowdowns-different-budget} where our PFA autoscaler shows better workload performance than the plan-based autoscalers.

%% file: sections/threats-to-validity.tex
\section{Threats to Validity}\label{ch5:sec:related-work}
The limitations of this study are mostly expressed in the number of considered
resource types and users.
Having two resource types is the minimum requirement
for comparing the performance of the considered dynamic and plan-based autoscalers.
A higher number of resource types would lead to higher planning complexity of
the plan-based autoscalers and would require to increase the total number of resources.
Adding more resources would also increase the WMS overhead, thus,
lengthening the experiment duration and bringing unnecessary complexity,
while not being relevant to the conceptual part of the paper.

The main reason for having multiple concurrent users serving independent workloads
is to model a background load present in real production clouds,
while conducting experiments in an isolated cluster environment.
The isolated cluster environment allows to achieve higher control
over the setup and the factors affecting the results.
Having only two users has the same reason as having two resource types---it
is sufficient for the initial performance comparison of the autoscalers.
We do not consider resource allocation and deallocation delays as
adding them would relatively equally affect all the considered autoscalers~\cite{ilyushkin2018experimental},
thus, making the effect of such delays not relevant to the autoscaler comparison purposes.

The choice of throughput as reliable metric can also be questioned.
However, in this work we are dealing with a workload that consists of many tasks with different durations,
and the scheduler that is used with PFA assigns tasks to the resources randomly,
so that each resource type processes tasks with different durations.
This allows for rather precise estimation of resource type speeds.
In the worst case, when there are no throughput information available,
PFA falls back to an equal share of resource types.
Using, for example, historical task runtimes we would need to suppose that
resources with shorter observed average task runtimes are faster.
In contrast with task runtimes, throughput allows to easier estimate
the resource speed on an autoscaling interval.

The goal of the considered autoscalers is not only to fit within
the budget but also to spend it effectively.
An infinite budget makes our problem equivalent to the cost minimization problem.
However, considering only cost minimization without budget constraint is not realistic as
cloud providers usually employ usage quotas and allow to set limits for monetary costs.

We have tried multiple parameters for MA and EWMA to find which of them reduce
the experienced workflow slowdowns. The considered parameter values can be used directly in practice,
however, other workload types may require additional parameter tuning.

%% file: sections/related-work.tex
\section{Related Work}\label{ch5:sec:related-work}

In this section, we overview specialized autoscaling policies for workflows
that focus on the resource-allocation problem.

{\bf State-of-the-art autoscaling policies:}
The Dynamic Scaling Consolidation Scheduling (DSCS)~\cite{mao2011auto},
Partitioned Balanced Time Scheduling (PBTS)~\cite{byun2011cost},
IaaS Cloud Partial Critical Paths (IC-PCP)~\cite{abrishami2013deadline},
Deadline Constrained Critical Path (DCCP)~\cite{arabnejad2017scheduling},
Dyna~\cite{zhou2016monetary}, and Partition Problem-based Dynamic Provisioning
and Scheduling (PPDPS)~\cite{singh2018novel} autoscalers
combine scheduling and allocation approaches, and, in contrast
to the approach used in this paper, have the goal
to minimize the operational cost under unlimited budget and meet
(soft) workflow deadlines. DSCS, PBTS, and Dyna are online plan-based autoscalers,
while IC-PCP, DCCP, and PPDPS are offline autoscalers.
The Plan autoscaler~\cite{ilyushkin2018experimental} is an online
plan-based autoscaler which does not support budget constraints and requires task runtime estimates
for the workflow tasks.
The Token autoscaler~\cite{ilyushkin2018experimental} is an online dynamic autoscaler
that uses a token-based approach to estimate the demand and
requires runtime estimate for the whole workflow.
The Dynamic Provisioning Dynamic Scheduling (DPDS)~\cite{malawski2015algorithms} is
an offline dynamic autoscaler for ensembles of scientific workflows
that supports a single resource type only. The autoscaler is threshold-based,
the cost- and deadline-constraints should be provided for the whole ensemble.
The Static Provisioning Static Scheduling (SPSS)~\cite{malawski2015algorithms} is
an offline autoscaler that creates a plan for each workflow in the ensemble,
and rejects workflows that exceed the deadline or budget.
BAGS~\cite{rodriguez2017budget} is a plan-based offline autoscaler
that partitions workflows into bags-of-tasks and
then applies a MIP-based approach to make the allocation plan.
The majority of the considered works performs simulations when evaluating the proposed algorithms.

{\bf Comprehensive comparisons and benchmarks:}
Versluis et al.~\cite{versluis2018trace} and Ilyushkin et al.~\cite{ilyushkin2018experimental}
perform comprehensive analysis
of different autoscalers for workloads of workflows.
Overall, these studies emphasize the need for autoscalers that can
cope with workloads of workflows,
but neither propose the autoscalers that support cost constraints and
multiple resource types, nor assess the time taken by autoscalers
to make decisions or evaluate the scalability. 
The recent survey~\cite{lu2018review} on cost
and makespan-aware workflow scheduling in cloud
provides a good overview of the current
scheduling and autoscaling trends for workflows.

%% file: sections/conclusions.tex
\section{Conclusions}\label{ch5:sec:conclusions}
We presented the novel Performance-Feedback Autoscaler (PFA)
for workloads of workflows.  
To make autoscaling decisions, PFA analyzes
historical task throughput and uses current workflow
structural information, instead of relying on task runtime estimates.
This makes PFA easier to use, as observing task throughput normally
requires less effort than obtaining task runtime estimates.

Overall, PFA has lower time-complexity and effectively 
minimizes workflow slowdowns, compared to two state-of-the-art
online plan-based autoscalers.
Our real-world experiments with the Apache Airflow workflow
management system show that PFA, compared to other two autoscalers,
has better applicability potential due to its good scalability
when dealing with possible demand surges, and good
end-user and system-oriented characteristics.

For future work, we plan to investigate the performance footprint of PFA
for other resource types, e.g., memory.
To further evaluate the scalability of the proposed autoscaler
we will consider setups with higher number of resource types and concurrent users.
To make PFA even more autonomous,
we will look how to automatically configure the signal smoothing
and try different feedback mechanisms.